\documentclass[preprint,authoryear,11pt]{elsarticle}

\usepackage[margin=1 in]{geometry}
\usepackage{amssymb}
\usepackage{amsmath}
\usepackage{xurl}
\usepackage{graphicx} 
\usepackage{subcaption}
\usepackage{multicol}
\usepackage{array}
\usepackage{mathrsfs}
\usepackage{multirow}
\usepackage[export]{adjustbox}
\usepackage{silence}
\usepackage{subcaption}
\usepackage{titlesec}
\usepackage{booktabs}
\usepackage{comment}
\usepackage[utf8]{inputenc}
\usepackage{nomencl}
\makenomenclature
\usepackage{etoolbox}
\usepackage{bm}
\newcommand{\mathbit}[1]{\bm{#1}}
\usepackage{enumitem}

\usepackage{caption}   
\usepackage{booktabs,makecell,tabularx,threeparttable,amssymb}
\usepackage{rotating}   

\usepackage{geometry}    
\usepackage{array}       
\usepackage{tabularx}    
\usepackage{booktabs}    
\usepackage{longtable}  
\usepackage[most]{tcolorbox}           
\usepackage{graphicx}      
\usepackage{needspace}     
\newcommand{\NomenRef}{\hyperref[nomenclature]{Table of Nomenclature}}

\usepackage{xcolor}          

\usepackage{enumitem}        

\usepackage[
  colorlinks=true,
  citecolor = blue,          
  urlcolor  = blue,          
  linkcolor = black          
]{hyperref}

\begin{document}

\begin{frontmatter}

\title{Designing a Multi-Period Model for Economic and Low-Carbon Hydrogen Transportation in Texas} 



\author[1]{Yixuan Huang}
\ead{yhuang77@cougarnet.uh.edu}
\cortext[cor1]{Corresponding Author}

\author[1]{Kailai Wang\corref{cor1}}
\ead{kwang43@central.uh.edu}

\author[2,3]{Jian Shi}
\ead{jshi14@uh.edu}

\affiliation[1]{organization={Department of Industrial and Systems Engineering, University of Houston},
            city={Houston},
            postcode={77204}, 
            state={Texas},
            country={United States}}

\affiliation[2]{organization={Department of Engineering Technology, University of Houston},
            city={Houston},
            postcode={77204}, 
            state={Texas},
            country={United States}}

\affiliation[3]{organization={Department of Electrical and Computer Engineering, University of Houston},
            city={Houston},
            postcode={77204}, 
            state={Texas},
            country={United States}}

\begin{abstract}
The transition to hydrogen powered transportation requires regionally tailored yet scalable infrastructure planning. This study presents the first Texas specific, multi-period mixed integer optimization model for hydrogen transportation from 2025 to 2050, addressing challenges in infrastructure phasing, asset coordination, and multimodal logistics. The framework introduces three innovations: (1) phased deployment with delayed investment constraints, (2) dynamic modeling of fleet aging and replacement, and (3) a clustering-based hub structure enabling adaptive two-stage hydrogen delivery. Simulations show pipeline deployment supports up to 94.8\% of hydrogen flow by 2050 under high demand, reducing transport costs by 23\% compared to vehicle-based systems. However, one-year construction delays reduce pipeline coverage by over 60\%, shifting reliance to costlier road transport. While the study focuses on Texas, its modular design and adaptable inputs apply to other regions. It provides a tool for policy makers and stakeholders to manage hydrogen transitions under logistical and economic constraints.

\end{abstract}




\end{frontmatter}


\section{Introduction}

As the global energy system shifts toward low-carbon solutions, hydrogen has emerged as a critical energy carrier with broad applications across industry, transportation, and power generation \citep{Zhang2024}. Among the various production pathways, green hydrogen produced via electrolysis using renewable electricity, has attracted increasing attention due to its significantly lower CO\textsubscript{2} emissions compared to fossil fuel based alternatives such as coal gasification and steam methane reforming \citep{KumarLim2022}. This environmental advantage is driving its expansion beyond traditional sectors like oil refining, coal chemicals, and steelmaking to emerging applications such as fuel cell vehicles (FCEV) and grid scale energy storage \citep{KumarLim2022, Vijayakumar2022}.

To realize hydrogen’s full potential as a sustainable fuel, particularly in transportation, it is essential to develop efficient and coordinated supply chain infrastructure. This includes refueling stations, distribution hubs, pipelines, and logistics systems that work in tandem. The design and operation of hydrogen transportation networks directly influence lifecycle costs and supply chain performance, making integrated planning a central challenge \citep{Li2021}. Yet despite growing momentum, hydrogen infrastructure development faces persistent technical, economic, and regulatory obstacles that constrain its scalability and long-term viability.

First, hydrogen infrastructure, such as pipelines, port terminals, and underground storage facilities, typically requires long development timelines. For example, a single hydrogen pipeline can take seven to twelve years to plan and construct due to complex permitting processes, regulatory coordination, and engineering demands \citep{PwC2023}. These lead times often exceed those for production plants or refueling stations, resulting in misaligned timelines across the supply chain. Second, infrastructure investment remains insufficient. Although global investment in electrolyzer technologies reached a record \$2.9billion in 2023, nearly five times the 2022 level, it represents only 6\% of the \$50billion in annual investment needed to achieve Net Zero Emissions targets by 2030 \citep{IEA2024}. This underinvestment underscores a substantial funding gap. Third, hydrogen transmission and distribution projects face planning and regulatory uncertainties. Delays caused by inconsistent demand forecasts, unclear production commitments, and fragmented regulatory frameworks have stalled major pipeline initiatives across the U.S. \citep{Bade2024}. These factors together create bottlenecks in supply chain development, while high costs and regulatory complexity continue to deter private-sector participation.

In this context, Texas represents both a challenge and an opportunity. The state is poised to become a key player in the national hydrogen economy, with a concentration of production facilities along the Gulf Coast and strong potential for integration with both renewable and natural gas resources. The Gulf Coast Hydrogen Hub, supported by up to \$1.2~billion in federal funding \citep{DoEHub2024}, aims to establish a backbone of low-carbon hydrogen supply for industrial and transportation sectors. However, Texas lags behind states like California in terms of hydrogen infrastructure for transportation. As of 2024, it lacks a single operational hydrogen refueling station \citep{ICCT2023}. In contrast, California’s extensive refueling network has facilitated the early adoption of hydrogen-powered vehicles.

To bridge this infrastructure gap, recent efforts are underway. The North Central Texas Council of Governments has received a \$70~million grant to construct up to five hydrogen refueling stations in key metropolitan areas, including Dallas–Fort Worth, Houston, Austin, and San Antonio \citep{Leonard2024}. These stations are designed primarily for medium and heavy duty trucks and are expected to anchor a hydrogen corridor linking Texas to Southern California. Given the strategic location of hydrogen production on the Gulf Coast, optimizing the state's hydrogen transportation network is essential to reduce distribution costs and facilitate widespread FCEV adoption. Effective planning must consider both regional demand variation and the need for scalable and cost-effective infrastructure deployment.

This study develops a multi-period, mixed-integer optimization model that simulates hydrogen transportation across Texas from 2025 to 2050. As the first geography-specific analysis focused on Texas, this work fills a critical gap by tailoring the modeling framework to the state’s unique spatial, economic, and infrastructural conditions. The model evaluates the optimal allocation of hydrogen delivery routes and transport modes from production facilities to refueling stations that incorporates phased infrastructure expansion, vehicle and pipeline lifecycle constraints, and multimodal logistics strategies. It further examines the role of strategically located hydrogen hubs in enhancing distribution efficiency within the Texas context. By balancing economic and environmental trade-offs over time, the model serves as a decision support tool to guide policymakers and industry stakeholders to plan a robust, regionally adaptive hydrogen transportation system tailored to Texas’s energy landscape.

The remainder of the paper is structured as follows. Section~2 reviews prior work and highlights key gaps that motivate the contributions of this study. Section~3 introduces the multi-period mixed-integer programming model, which optimizes hydrogen transportation under two modes: direct delivery from production sites to refueling stations and a hub-based distribution strategy. The section details the objective function, constraints, and incorporation of hydrogen hubs to improve cost efficiency. Section~4 presents a case study in Texas, discussing data collection, demand forecasting, and scenario construction. It then examines the model’s outputs, assessing the impact of different supply chain configurations and infrastructure planning strategies on hydrogen distribution and optimized network performance. Section~5 concludes the study with key findings and policy recommendations for hydrogen infrastructure planning in Texas. It also discusses limitations that could impact the implementation of the proposed transportation strategies.

\section{Literature Review}

Research on hydrogen supply chain (HSC) transportation has evolved along three main methodological lines: techno-economic analysis (TEA), mixed integer programming (MIP)-based optimization, and alternative or hybrid approaches focused on uncertainty and environmental performance. 

TEA is frequently employed to assess the economic viability and cost structure of HSC systems. For example, \citet{Perna2023} investigated maritime hydrogen transportation networks, emphasizing the infrastructure requirements for production, storage, and distribution. \citet{Chen2021} used HOMER Pro to compare onsite and offsite hydrogen production strategies at a refueling station in Shanghai, optimizing for net present cost and levelized cost of hydrogen. Extending this work, \citet{Yu2024} developed a techno-economic model for hydrogen pipeline systems in China, analyzing how design pressure, flow rate, and distance affect transportation cost. \citet{Rong2024} provided a comparative evaluation of multiple transport modes which includes compressed gas hydrogen, liquid hydrogen, pipeline hydrogen, and LOHC to identify cost optimal storage and transport configurations. These TEA studies contribute to understanding cost drivers and technology selection across the HSC. {\textit{Research Gap:} Existing TEA efforts are typically static, as they rely on predefined system configurations and fixed input assumptions, which limits their ability to capture dynamic factors related to scalability of hydrogen systems over time, such as phased infrastructure deployment, evolving demand, technology learning curves, or time dependent policy and market conditions.}

Complementing these TEA efforts, a growing body of research has utilized MIP-based optimization models to plan HSC infrastructure and logistics. \citet{Forghani2023} modeled hydrogen pipeline network selection across multiple planning periods, but did not incorporate phased construction or infrastructure evolution. \citet{Zhao2022} added transportation mode flexibility to the modeling framework, yet omitted pipeline lifecycle and capacity constraints. \citet{Yoon2022} introduced natural gas pipeline repurposing and truck-based transport options to improve system flexibility. \citet{Feng2024} focused on centralized storage facilities and their role in balancing regional supply and demand, while \citet{Sizaire2024} and \citet{Dautel2024} embedded decarbonization targets and renewable energy supply variability into the optimization framework. These models offer powerful tools to minimize system cost under technical constraints. {\textit{Research Gap:} 
Most existing optimization models are limited to single period scopes, and do not account for phased construction, fleet lifecycle dynamics, and spatially detailed logistics, which constrains their applicability to long term and region specific hydrogen infrastructure planning.}

Beyond TEA and MIP, several studies adopt alternative approaches to account for uncertainty and environmental performance. \citet{Deng2023} used distributionally robust optimization to manage hydrogen replenishment in response to demand volatility and pipeline disruptions, allowing flexible transitions between truck and pipeline based delivery. Meanwhile, \citet{Hermesmann2023} applied life cycle assessment (LCA) to compare emissions trade-offs between local production and long distance hydrogen transport, which considers renewable energy availability and transport distance. These studies expand the scope of analysis to include climate impact and system resilience. {\textit{Research Gap:} Existing approaches remain largely disconnected from detailed infrastructure deployment frameworks that limit their ability to inform coordinated investment timing, operational scheduling, and the sequential integration of transport modes and technologies over the system’s lifespan.}

Another underdeveloped area in the literature concerns the treatment of hydrogen hubs. While \citet{Saldanha2022} provides a systematic review of hub location problems in general logistics contexts, their application to hydrogen networks remains limited. \citet{Yoon2022} and \citet{Feng2024} treated hubs as endogenous storage or production nodes whose locations are optimized within the model. However, these approaches do not explicitly define hubs as regional redistribution centers, nor do they impose constraints on upstream versus downstream transport modes. {\textit{Research Gap:} Existing hydrogen hub models do not fully capture the transshipment dynamics, delivery boundaries, or operational clarity required for large scale HSC planning in spatially heterogeneous regions.}

To address these research gaps, this study develops a multi-period, mixed-integer optimization framework that integrates infrastructure phasing, lifecycle management, and spatially grounded hub structuring {Table~\ref{tab:design_features} summarizes recent multi-period hydrogen transportation optimization models and highlights their design characteristics. While prior studies capture individual aspects such as transport mode flexibility, carbon emissions, or pipeline planning, few address the joint coordination of phased infrastructure rollout, asset lifecycle management, and regionalized hub-based routing. In contrast, this study uniquely integrates these elements, tailored specifically to the hydrogen infrastructure and geographic clustering present in Texas. The main contributions of this study can be summarized as:} 

\begin{table}[t!]
    \centering
    \caption{{\textit{Comparison of Design Features Across Multi-Period Hydrogen Transportation Models in the Literature}}}
    \includegraphics[width=1.0\textwidth]{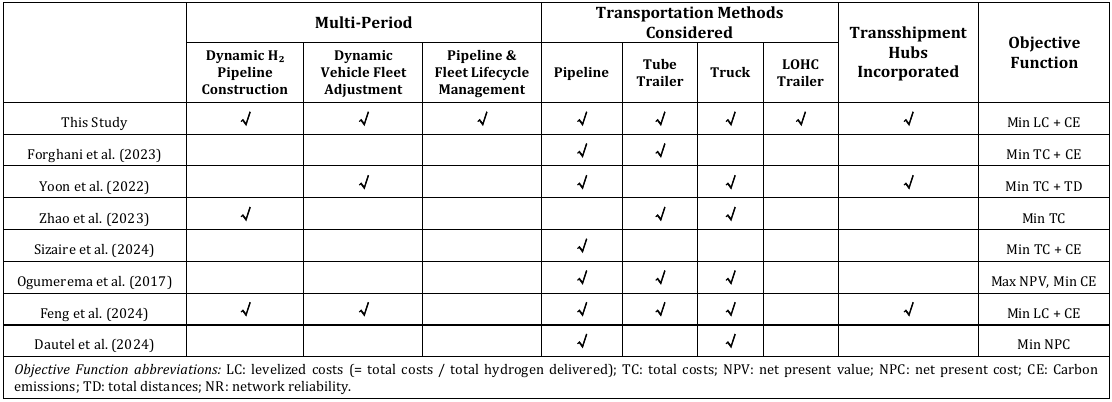}
    \label{tab:design_features}
\end{table}

\vspace{-4mm}
\begin{enumerate}[label=\arabic*)]

\item \textbf{Phased Infrastructure Deployment with Time-Delayed Investment:} The proposed model captures the temporal complexity of hydrogen infrastructure development through the incorporation of phased construction schedules, annual deployment limits, and constraints on concurrent projects. By accounting for construction cycle duration and resource limitations, it reflects the long lead times and sequencing challenges observed in real world hydrogen infrastructure, particularly in large-scale, multi-jurisdictional contexts like Texas.\vspace{-2mm}

\item \textbf{Fleet and Pipeline Lifecycle Management:} The proposed model integrates detailed vehicle and pipeline lifecycle dynamics that cover procurement, aging, and retirement of hydrogen delivery fleets, as well as expansion and replacement of pipeline infrastructure. This enables joint optimization of long-term operational sustainability, asset utilization, and cost efficiency, which are often overlooked in single-period or static infrastructure models. \vspace{-2mm}

\item \textbf{Cluster-Driven Hydrogen Hubs with Two-Stage Transport Structuring:} The proposed model introduces a clustering-based method to define hydrogen hubs exogenously as fixed regional transshipment zones for enabling spatially structured and region specific network design. Unlike prior studies, this approach uniquely tailors the hub-and-spoke transport structure to the distinct infrastructure, regulatory, and market conditions of Texas, which makes it the first of its kind to model hydrogen distribution in Texas at this level of detail and specificity. \vspace{-2mm}

\end{enumerate}
\color{black}

\section{Methodology}
\subsection{Problem Statement and Model Description}
\begin{figure}[h!]
    \centering
     \includegraphics[scale=0.7]{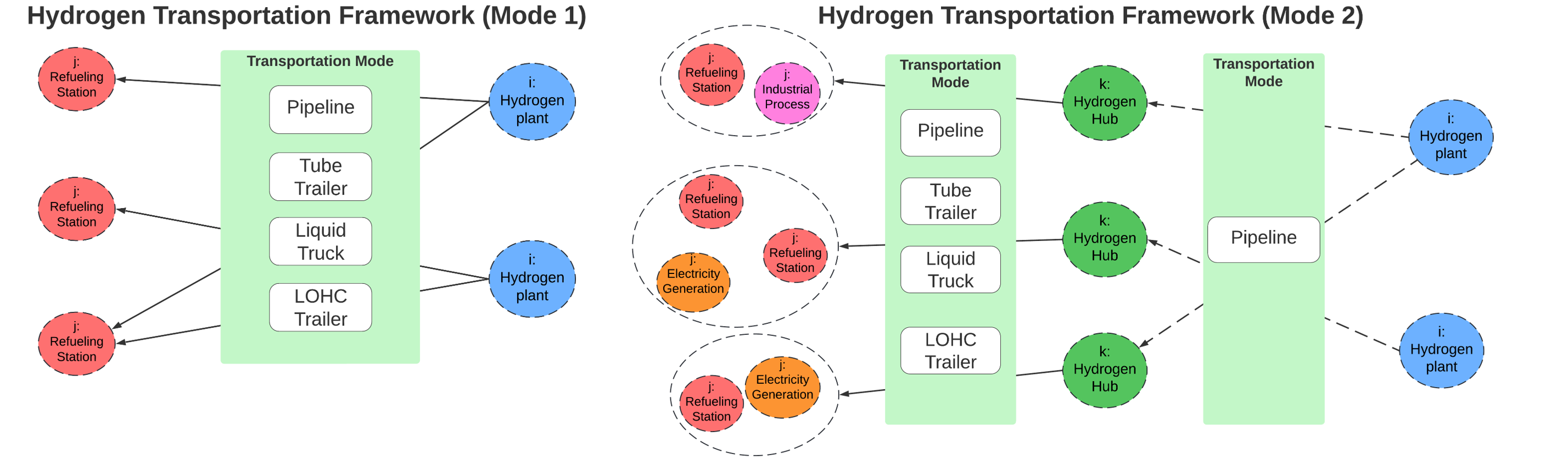}  
     \caption{\textit{Hydrogen Transportation Framework}}
    \label{figure1}
\end{figure}

This study adopts a multi-period mixed-integer programming model to optimize hydrogen transportation networks from plants to refueling stations, aiming to minimize the overall levelized costs associated with economic expenditures and carbon emissions.  The hydrogen supply chain structures shown in \autoref{figure1} initially focus on four transportation methods: pipeline, tube trailer, liquid truck, and LOHC trailer. This basic framework facilitates modeling a more economical and environmentally friendly distribution of transport methods over various periods while meeting the hydrogen demand requirements.

In \textbf{Mode 1}, the primary objective is direct transportation from hydrogen production plants to refueling stations. The distribution of hydrogen is straightforward and efficient for idealized, simplified scenarios; however, it does not account for the complexities of intermediate hubs. 

Building on this foundation, \textbf{Mode 2} introduces the concept of the hydrogen hub, which primarily supplies hydrogen fuel for transportation while also redistributing surplus hydrogen.The main goal remains fueling hydrogen vehicles, but hubs provide flexibility by redirecting excess hydrogen, which was originally allocated for refueling stations, to electricity-generation sites or industrial processes when needed. This ensures efficient utilization of hydrogen resources, supplementing regional shortages in other sectors. In Mode 2, the first stage, $i \rightarrow k$, strictly relies on pipeline transportation, delivering bulk hydrogen to hubs, while the second stage, $k \rightarrow j$, offers flexibility by allowing multiple transport modes, including pipelines and vehicles. This structure reduces long-distance vehicle transport costs while leveraging pipelines for high capacity delivery, a design choice influenced by previous studies, where \citet{Yoon2022} used repurposed natural gas pipelines for regional transport while trucks and predefined pipelines handled local distribution. Likewise, \citet{Dogliani2024} only incorporated pipelines and ships for bulk hydrogen transmission. This design reinforces the efficiency of pipelines for long-distance hydrogen transport and supports flexible final stage distribution. 

Below we will first focus on the modeling of Mode 1 and describe how hydrogen transportation is optimized directly from plants to demand nodes. Then we will examine the formulation of Mode 2 and provide a detailed explanation of the incorporation of hydrogen hubs.

The main assumptions of this study are as follows, ensuring the model remains rational and applicable:  
\begin{enumerate}[label=\arabic*)]
  \item The budget is adequate but limited, meaning that while the necessary infrastructure can be afforded, cost efficiency remains a key consideration.  
  \item All model parameters are assumed to be known or, in cases where data are unavailable, are assigned reasonable estimates based on industry standards and prior research.  
  \item The distance between any two points is approximated as a straight-line distance, without accounting for potential geographical or infrastructural obstacles.  
  \item Within each transportation mode, no significant variation in cost, efficiency, or material composition is considered; for example, all pipelines are assumed to have uniform installation costs and material properties, and all liquid trucks operate under the same performance parameters.  
  \item Hydrogen production sites do not generate local demand, and demand sites do not have local hydrogen production, ensuring that all hydrogen must be transported between distinct supply and demand points.  
\end{enumerate}

The summary of the variables, parameters, and sets used in this study is provided in the \NomenRef{}. Each variable and parameter is defined with its specific role in the hydrogen supply-chain optimization framework, including costs from various aspects, as well as constraints related to mass balance, pipeline construction, vehicle fleet management, and environmental regulations.
\begin{center}
\small                                    
{%
  \begin{longtable}{@{}p{0.30\linewidth}p{0.67\linewidth}@{}}
  \caption*{Nomenclature}\label{nomenclature}\\
  \toprule
  \multicolumn{2}{@{}l}{\textbf{Sets}}\\
  \midrule
  $i \in I$ & Set of hydrogen supply nodes\\
  $j \in J$ & Set of hydrogen demand nodes\\
  $r \in R$ & Set of transportation methods (1 Pipeline; 2 Tube trailer; 3 Liquid truck; 4 Liquid Organic Hydrogen Carrier (LOHC) trailer)\\
  $t \in T$ & Set of time periods\\
  $k \in K$ & Set of hub nodes\\[6pt]

  \multicolumn{2}{@{}l}{\textbf{Parameters}}\\
  \addlinespace[2pt]
  \multicolumn{2}{@{}l}{\emph{Financial Parameters}}\\
  \midrule
  $\beta$ & Discount rate\\
  $C_{t}^{KH}$ & Unit capital cost for H\textsubscript{2} pipeline built (\$/km)\\
  $C_{rt}^{KV}$ & Unit capital cost for new transportation vehicle (\$/vehicle)\\
  $P_{rt}^{\mathrm{fuel}}$ & Fuel price for transportation method $r$ at period $t$ (\$/Liter)\\
  $P_{rt}^{\mathrm{wage}}$ & Driver wage for transportation method $r$ at period $t$ (\$/hour)\\
  $P_{t}^{H}$ & Unit pipeline maintenance fee (\$/km)\\
  $\alpha$ & Unit penalty cost of hydrogen loss (\$/kg)\\
  $w_{r}^{\mathrm{loss}}$ & Hydrogen loss per transported unit (kg/km)\\
  $\delta$ & Unit penalty cost of CO\textsubscript{2} emission (\$/kg)\\
  $em_{r}$ & Unit amount of CO\textsubscript{2} emission (kg CO\textsubscript{2}\,/kg H\textsubscript{2} transported)\\
  $\lambda$ & Penalty cost for surplus or shortage (\$/kg)\\
  $M$ & A sufficiently large constant used in constraints to enforce logical implications\\
  $C_{ikt}^{KH},\,C_{kjt}^{KH}$ & Unit capital cost for H\textsubscript{2} pipeline built ($i{\to}k$, $k{\to}j$) (\$/km)\\[4pt]

  \multicolumn{2}{@{}l}{\emph{Physical \& Technical Parameters}}\\
  \midrule
  $L_{ij}$ & Shortest distance between node $i$ and $j$ (km)\\
  $Q_{it}^{S}$ & Maximum hydrogen supply capacity at supply node $i$ during period $t$ (kg/period)\\
  $Q_{jt}^{D}$ & Hydrogen demand at demand node $j$ during period $t$ (kg/period)\\
  $FE_{r}$ & Fuel economy of transportation method $r$ (km/Liter)\\
  $V_{r}^{\mathrm{cap}}$ & Load capacity of transportation method $r$ (kg/trip)\\
  $SP_{r}$ & Average speed of transportation method $r$ (km/h)\\
  $LT_{r}$ & Loading/unloading time of transportation method $r$ (hours)\\
  $AT_{r}$ & Availability time of transportation method $r$ (hours/day)\\
  $T_{r}^{\max}$ & Maximum lifespan of transportation method $r$ (in years)\\
  $EM_{jt}^{\mathrm{ceil}}$ & Maximum allowable CO\textsubscript{2} emissions at node $j$ in period $t$ (kg)\\
  $\mathrm{NH}_{\max}$ & Maximum number of pipelines that can be built in one year\\
  $\mathrm{NH}_{\mathrm{gap}}$ & Construction period for pipelines (in years)\\
  $F_{\mathrm{base}}^{\max}$ & Unit length pipeline base flow limit (kg/km/period)\\
  $L_{\mathrm{base}}^{H}$ & Reference length for pipeline flow calculations (km)\\[4pt]

  \multicolumn{2}{@{}l}{\emph{Clustering-Based Parameter}}\\
  \midrule
  $c_{j}$ & The assigned hub index for demand node $j$. In other words, each demand node $j\in J$ is uniquely mapped to exactly one hub node $k\in K$ through clustering. Formally, if $c_{j}=k$, then demand node $j$ belongs to the cluster associated with hub $k$.\\[6pt]

  \multicolumn{2}{@{}l}{\textbf{Variables}}\\
  \addlinespace[2pt]
  \multicolumn{2}{@{}l}{\emph{Intermediate Variables}}\\
  \midrule
  $PV_t$ & Present value factor at time $t$, calculated as: $PV_t=\dfrac{1}{(1+\beta)^t}$\\
  $S_{jt}$ & Surplus or shortage of hydrogen at demand point $j$ at time $t$ (kg), calculated as: $S_{jt}=S_{jt}^{\mathrm{pos}}-S_{jt}^{\mathrm{neg}}$\\[4pt]

  \multicolumn{2}{@{}l}{\emph{Integer Variables}}\\
  \midrule
  $NV_{rt}$ & Number of transportation vehicles of type $r$ at time $t$\\
  $NV_{rt}^{\mathrm{new}}$ & Number of newly built transportation vehicles of type $r$ at time $t$\\
  $NV_{rt}^{\mathrm{scrap}}$ & Number of scrapped transportation vehicles of type $r$ at time $t$\\[4pt]

  \multicolumn{2}{@{}l}{\emph{Continuous Variables}}\\
  \midrule
  $V_{ijrt}$ & Amount of hydrogen transported from node $i$ to node $j$ via transportation method $r$ at time $t$ (kg)\\
  $S_{jt}^{\mathrm{pos}}$ & Surplus of hydrogen at demand point $j$ at time $t$ (kg)\\
  $S_{jt}^{\mathrm{neg}}$ & Shortage of hydrogen at demand point $j$ at time $t$ (kg)\\
  $V_{ikt}$ & Amount of hydrogen transported from node $i$ to node $k$ at time $t$ (kg)\\
  $V_{kjrt}$ & Amount of hydrogen transported from node $k$ to node $j$ via transportation method $r$ at time $t$ (kg)\\[4pt]

  \multicolumn{2}{@{}l}{\emph{Binary Variables}}\\
  \midrule
  $BH_{ijt}$ & Binary variable indicating whether an H\textsubscript{2} pipeline exists from node $i$ to $j$ at time $t$ (1 if exists, 0 otherwise)\\
  $BH_{ijt}^{\mathrm{new}}$ & Binary variable indicating whether a new H\textsubscript{2} pipeline is constructed from node $i$ to $j$ at time $t$ (1 if constructed, 0 otherwise)\\
  $BH_{ikt}$ & Binary variable indicating whether an H\textsubscript{2} pipeline exists from node $i$ to $k$ at time $t$ (1 if exists, 0 otherwise)\\
  $BH_{kjt}$ & Binary variable indicating whether an H\textsubscript{2} pipeline exists from node $k$ to $j$ at time $t$ (1 if exists, 0 otherwise)\\
  $BH_{ikt}^{\mathrm{new}}$ & Binary variable indicating whether a new H\textsubscript{2} pipeline is constructed from node $i$ to $k$ at time $t$ (1 if constructed, 0 otherwise)\\
  $BH_{kjt}^{\mathrm{new}}$ & Binary variable indicating whether a new H\textsubscript{2} pipeline is constructed from node $k$ to $j$ at time $t$ (1 if constructed, 0 otherwise)\\
  \bottomrule
  \end{longtable}
  \addtocounter{table}{-1} 
}
\end{center}

\subsection{Objective Function}\label{subsec:objective}

The objective is to minimize the levelized cost of hydrogen transportation over the whole-time horizon, $\mathrm{LCH}_2$, which is derived similarly to the levelized cost of energy (LCOE) approach, as defined by \citet{Ganter2024}. It is obtained by dividing the total net present cost (TC) of the supply chain by the total net present hydrogen delivered (TV), ensuring an economically comparable assessment of transportation strategies:
\begin{equation}\label{eq:LCH2}
  \min \mathrm{LCH}_2
  = \frac{\text{TC}}{\text{TV}}
  =\frac{\text{CC} + \text{MC} + \text{OC}}{\sum_{i,j,r,t} V_{ijrt}}
\end{equation}

\noindent where the total hydrogen delivered is calculated by adding up the transportation volume of various methods between regions in each period; the total system cost consists of capital cost (CC), maintenance and operational cost (MC), and other cost(OC).

\textbf{The capital cost} accounts for investments in hydrogen transportation infrastructure, specifically hydrogen pipeline construction ($\mathbit{C}_{\mathbit{HC}}$) and the purchase cost of hydrogen transport vehicles ($\mathbit{C}_{\mathbit{VC}}$):
\begin{equation}\label{eq:CapCost}
  \text{CC}
  =  \mathbit{C}_{\mathbit{HC}}
  +  \mathbit{C}_{\mathbit{VC}}
\end{equation}
\begin{equation}\label{eq:HCC}
  \mathbit{C}_{\mathbit{HC}}
  = \sum_{t}\mathrm{PV}_{t}\sum_{i,j}
    B\!H_{ijt}^{\mathrm{new}}\,
    C_{ijt}^{KH}\,L_{ij}
\end{equation}
\begin{equation}\label{eq:VCC}
  \mathbit{C}_{\mathbit{VC}}
  = \sum_{t}\mathrm{PV}_{t}\sum_{r\neq1}
    C_{rt}^{KV}\,N\!V_{rt}^{\mathrm{new}}
\end{equation}

\noindent where $\mathbit{i}\in\mathbit{I}$, $\mathbit{j}\in\mathbit{J}$, $\mathbit{r}\in\mathbit{R}$, $\mathbit{t}\in\mathbit{T}$ represents hydrogen supply points (hydrogen plants), the hydrogen demand points (refueling stations), transportation methods (\#1 pipeline, \#2 tube trailer, \#3 liquid truck, and \#4 LOHC trailer), and time periods (in years), respectively; $\mathbit{L}_{\mathbit{ij}}$ denotes the distance between $\mathbit{i}$ and $\mathbit{j}$. $\mathbit{N}\mathbit{V}_{\mathbit{rt}}^{\mathrm{new}}$ represents the number of newly purchased vehicles, while $\mathbit{C}_{\mathbit{ijt}}^{\mathbit{KH}}$ and $\mathbit{C}_{\mathbit{rt}}^{\mathbit{KV}}$ are the unit capital costs for building hydrogen pipelines and purchasing vehicles, respectively. $\mathbit{B}\mathbit{H}_{\mathbit{ijt}}^{\mathrm{new}}$ is a binary variable, where a value of 1 indicates that one new hydrogen pipeline is constructed between $\mathbit{i}$ and $\mathbit{j}$, and 0 otherwise. The present value factor $\mathbit{P}\mathbit{V}_{\mathbit{t}}$ is defined as $\mathbit{P}\mathbit{V}_{\mathbit{t}}=\dfrac{\mathbf{1}}{\left(\mathbf{1}+\mathbf{\beta}\right)^{\mathbit{t}}}$ \citep{Arnaboldi2014}.

\textbf{The maintenance and operational cost} is decomposed into fuel cost ($\mathbit{C}_{\mathbit{FO}}$), driver’s labor cost ($\mathbit{C}_{\mathbit{LO}}$) for vehicle transportation, and the pipeline maintenance cost ($\mathbit{C}_{\mathbit{HO}}$). The formulations for $\mathbit{C}_{\mathbit{FO}}$ and $\mathbit{C}_{\mathbit{LO}}$ are adapted from \citet{Feng2024}, which incorporate detailed cost components for hydrogen vehicle transportation:
\begin{equation}\label{eq:OpCost}
  \text{MC}
  = \mathbit{C}_{\mathbit{FO}}
  + \mathbit{C}_{\mathbit{LO}}
  + \mathbit{C}_{\mathbit{HO}}
\end{equation}
\begin{equation}\label{eq:FOC}
  \mathbit{C}_{\mathbit{FO}}
  = \sum_{t}\mathrm{PV}_{t}\sum_{i,j}\sum_{r\neq1}
    P_{rt}^{\mathrm{fuel}}
    \frac{2\,L_{ij}\,V_{ijrt}}{F\!E_{r}\,V_{r}^{\mathrm{cap}}}
\end{equation}
\begin{equation}\label{eq:LOC}
  \mathbit{C}_{\mathbit{LO}}
  = \sum_{t}\mathrm{PV}_{t}\sum_{i,j}\sum_{r\neq1}
    P_{rt}^{\mathrm{wage}}
    \frac{2\,L_{ij}/S\!P_{r}+L\!T_{r}}{V_{r}^{\mathrm{cap}}}
    V_{ijrt}
\end{equation}
\begin{equation}\label{eq:HOC}
\mathbit{C}_{\mathbit{HO}}
  = \sum_{t}\mathrm{PV}_{t}\sum_{i,j}
    B\!H_{ijt}\,P_{t}^{H}\,L_{ij}
\end{equation}

\noindent where $\mathbit{B}\mathbit{H}_{\mathbit{ijt}}$ is a binary variable, with 1 indicating that there exists hydrogen pipeline between $\mathbit{i}$ and $\mathbit{j}$, 0 otherwise; the financial parameters $\mathbit{P}_{\mathbit{rt}}^{\mathbit{fuel}}$, $\mathbit{P}_{\mathbit{rt}}^{\mathbit{wage}}$, and $\mathbit{P}_{\mathbit{t}}^{\mathbit{H}}$ are unit price for fuel, driver wage, and pipeline maintenance; the physical parameters $\mathbit{F}\mathbit{E}_{\mathbit{r}}$, $\mathbit{V}_{\mathbit{r}}^{\mathbit{cap}}$, $\mathbit{S}\mathbit{P}_{\mathbit{r}}$, and $\mathbit{L}\mathbit{T}_{\mathbit{r}}$ denote the fuel economy, load capacity, vehicle average speed and loading/unloading time, respectively.

\textbf{The other cost} includes hydrogen loss cost ($\mathbit{C}_{\mathbit{HL}}$), carbon disposal cost ($\mathbit{C}_{\mathbit{CL}}$), and supply-demand imbalance cost ($\mathbit{C}_{\mathbit{SL}}$). Specifically, $\mathbit{C}_{\mathbit{SL}}$ introduces adaptive penalties to manage supply-demand mismatches, enabling redistribution to mitigate hydrogen shortages or surpluses:
\begin{equation}\label{eq:OtherCost}
  \text{OC}
  = \mathbit{C}_{\mathbit{HL}}
  + \mathbit{C}_{\mathbit{CL}}
  + \mathbit{C}_{\mathbit{SL}}
\end{equation}
\begin{equation}\label{eq:HLC}
  \mathbit{C}_{\mathbit{HL}}
  = \sum_{t}\mathrm{PV}_{t}\,
    \alpha
    \sum_{i,j,r}
    w_{r}^{\mathrm{loss}}\,
    L_{ij}\,
    \frac{V_{ijrt}}{V_{r}^{\mathrm{cap}}}
\end{equation}
\begin{equation}\label{eq:CLC}
\mathbit{C}_{\mathbit{CL}}
      =\sum_{t}\mathrm{PV}_{t}\,\delta
       \sum_{i,j,r} e m_{r}\,L_{ij}\,
       \frac{N\!V_{rt}}{F\!E_{r}},\quad\forall r\neq1
\end{equation}
\begin{equation}\label{eq:SLC}
\mathbit{C}_{\mathbit{SL}}
      =\sum_{t}\mathrm{PV}_{t}\,\lambda
       \sum_{j}\bigl(S_{jt}^{\mathrm{pos}}+S_{jt}^{\mathrm{neg}}\bigr)
\end{equation}

\noindent where $\mathbit{\alpha}$, $\mathbit{\delta}$, $\mathbit{\lambda}$ are penalty parameters for hydrogen loss, carbon emissions, and supply–demand mismatches, respectively; $\mathbit{w}_{\mathbit{r}}^{\mathbit{loss}}$ means the fraction of hydrogen lost per unit transported; $\mathbit{e}\mathbit{m}_{\mathbit{r}}$ means carbon emissions per unit of hydrogen transported.

\subsection{Constraints}\label{subsec:constraints}

\subsubsection{Mass Balance Constraints}

The transportation system ensures that hydrogen supply from production nodes meets the demand at consumption points while accounting for transportation losses. The system's mass balance is maintained by ensuring supply matches the transport volume, which in turn meets demand, achieving overall consistency and efficiency.

The \textbf{production supply capacity limit} (\ref{eq:SupplyCap}) ensures that the supply of each hydrogen plant $i$ does not exceed its production capacity $Q_{it}^{S}$ during any period $t$.
\begin{equation}\label{eq:SupplyCap}
\sum_{j,r} V_{ijrt} \le Q_{it}^{S},\quad \forall i,t
\end{equation}

\textbf{Demand satisfaction constraint} (\ref{eq:DemandSat}) ensures that the total hydrogen delivered to a demand node $j$ meets the required demand $Q_{jt}^{D}$. It also accounts for transportation losses during delivery and any surplus or shortage ($S_{jt}$) at the demand node:
\begin{equation}\label{eq:DemandSat}
\sum_{i,r}V_{ijrt}-w_{r}^{\mathrm{loss}}\!\cdot\!L_{ij}\,
\frac{V_{ijrt}}{V_{r}^{\mathrm{cap}}}
      =Q_{jt}^{D}+S_{jt},\quad \forall j,t
\end{equation}

\noindent where the surplus/shortage variable $\mathbit{S}_{\mathbit{jt}}$ is the net result of two components $\mathbit{S}_{\mathbit{jt}}^{\mathrm{pos}}$ and $\mathbit{S}_{\mathbit{jt}}^{\mathrm{neg}}$, representing surplus and shortage in hydrogen supply at the demand node:
\begin{equation}\label{eq:SNet}
S_{jt}=S_{jt}^{\mathrm{pos}}-S_{jt}^{\mathrm{neg}},\quad \forall j,t
\end{equation}

\noindent and where constraints (\ref{eq:Spos}) and (\ref{eq:Sneg}) enforce non-negativity on $S_{jt}^{\mathrm{pos}}$ and $S_{jt}^{\mathrm{neg}}$ to ensure that the surplus and shortage values are physically meaningful and non-negative, in line with the model's assumptions:
\begin{align}
S_{jt}^{\mathrm{pos}} &\ge 0,\quad \forall j,t\label{eq:Spos}\\
S_{jt}^{\mathrm{neg}} &\ge 0,\quad \forall j,t\label{eq:Sneg}
\end{align}

\subsubsection{Pipeline Constraints}

Pipeline constraints include construction constraints, operational requirements, and lifetime management, ensuring the number of pipelines under construction remains within resource and capacity constraints, regulating the conditions under which pipelines can be used, and managing their operational lifetimes in line with system planning. Details are outlined below.

\textbf{Pipeline Usage Restriction} ensures that transportation via pipelines is only allowed if the pipeline exists, i.e.\ $B\!H_{ijt}=1$. To enforce this condition, a sufficiently large constant $M$ is introduced, inspired by the big-$M$ formulation approach in \citet{Zhao2022}:
\begin{equation}
  V_{ijrt}\;\le\; M\cdot B\!H_{ijt},
  \quad r=1,\;\forall i,j,t
  \label{eq:PipeUse}
\end{equation}

\textbf{Pipeline construction system limit} ensures that the maximum number of pipelines under construction during the same period is limited to a threshold $\mathrm{NH}_{\max}$, reflecting the limitation on manpower and material resources for pipeline construction:
\begin{equation}
  \sum_{i,j} B\!H_{ijt}^{\mathrm{new}} \;\le\; \mathrm{NH}_{\max},
  \quad \forall t
  \label{eq:SysLimit}
\end{equation}

\textbf{Pipeline construction dynamics} follows two different progressions, a normal case requiring multiple time periods and a special case allowing instant completion when upgrading existing infrastructure:

\noindent \textit{1) Normal Case:} pipeline construction requires multiple time periods
($\mathrm{NH}_{\mathrm{gap}}>\mathbf{0}$)

Initially, all pipelines are absent at $t=0$:
\begin{equation}
  B\!H_{ij0}=0,\quad \forall i,j
  \label{eq:Init0}
\end{equation}

During construction, the status of pipelines remains unchanged due to the construction delay, representing the time gap between the start of construction and the pipeline becoming operational:
\begin{equation}
  B\!H_{ij\tau}=B\!H_{ij,\tau-1},
  \quad \forall i,j,\;
  t\le \tau < t+\mathrm{NH}_{\mathrm{gap}}
  \label{eq:Delay}
\end{equation}

Once construction is completed, the pipeline status is updated to reflect the reality of becoming operational:
\begin{equation}
  B\!H_{ij,t+\mathrm{NH}_{\mathrm{gap}}}
  =B\!H_{ij,t+\mathrm{NH}_{\mathrm{gap}}-1}
  +B\!H_{ijt}^{\mathrm{new}},
  \quad \forall i,j,t
  \label{eq:Update}
\end{equation}

Additionally, construction is prohibited if the pipeline cannot be completed within the planning horizon:
\begin{equation}
  B\!H_{ijt}^{\mathrm{new}} = 0,
  \quad \forall i,j,t~\text{where } t+\mathrm{NH}_{\mathrm{gap}}>T
  \label{eq:NoLateBuild}
\end{equation}

\noindent \textit{2) Special Case:} an idealized situation of instant pipeline construction
($\mathrm{NH}_{\mathrm{gap}}=\mathbf{0}$)

An applicable situation is directly upgrading natural gas pipelines to hydrogen pipelines, which can be quickly and easily modified for hydrogen transportation. In this case, pipelines become operational in the same time period they are constructed, with the initial and subsequent states of the pipeline being updated dynamically:
\begin{equation}
B\!H_{ijt}=
\begin{cases}
  B\!H_{ij0}^{\mathrm{new}}, & t=0,\\[4pt]
  B\!H_{ij,t-1}+B\!H_{ijt}^{\mathrm{new}}, & t\ge 1,
\end{cases}
\quad \forall i,j
\label{eq:InstantBuild}
\end{equation}

\textbf{Pipeline lifespan management} considers that pipelines are automatically decommissioned when their lifespan
$T_{r=1}^{\max}$ expires:
\begin{equation}
  B\!H_{ij,t+T_{r=1}^{\max}}
  \;\le\;
  1-B\!H_{ijt}^{\mathrm{new}},
  \quad \forall i,j,t~\text{where } t+T_{r=1}^{\max}\le T
  \label{eq:Lifespan}
\end{equation}

\textbf{Pipeline flow capacity} is inversely proportional to its length, accounting for pressure drops and frictional losses:
\begin{equation}\label{eq:PipeCap}
  V_{ijrt}\;\le\;
  F_{\mathrm{base}}^{\max}\,
  \frac{L_{\mathrm{base}}^{H}}{L_{ij}},
  \quad r=1,\;\forall i,j,t
\end{equation}

\subsubsection{Vehicle Constraints}

The vehicle constraints ensure that the hydrogen transportation demand is met by appropriately acquiring vehicles in each period, which are mainly categorized as follows.

\textbf{The number of vehicles required for transportation} is determined by the total hydrogen demand, travel distance, and operational constraints (speed, loading/unloading time) \citep{Feng2024}:
\begin{equation}\label{eq:MinVeh}
  N\!V_{rt}\;\ge\;
  \sum_{i,j}
  \frac{V_{ijrt}}{A\!T_{r}\,V_{r}^{\mathrm{cap}}}
  \frac{2L_{ij}}{S\!P_{r}+L\!T_{r}}
  \cdot  \frac{1}{365},
  \quad \forall r\neq1
\end{equation}

\textbf{The dynamic management of vehicles} that includes purchase, updates, and scrapping, is governed through the following process:

At the initial period, the number of vehicles for each vehicle transportation method $r$ is equal to the number of newly purchased vehicles:
\begin{equation}\label{eq:VehInit}
  N\!V_{r0}=N\!V_{r0}^{\mathrm{new}},\quad \forall r\neq1
\end{equation}

Dynamically, the fleet size is updated each period by adding newly purchased vehicles, retaining those from the previous period, and removing scrapped vehicles. Constraint (\ref{eq:VehDyn}) ensures the fleet reflects operational availability, which is defined as:
\begin{equation}\label{eq:VehDyn}
  N\!V_{rt}=N\!V_{rt}^{\mathrm{new}}
            +N\!V_{r,t-1}
            -N\!V_{rt}^{\mathrm{scrap}},
  \quad \forall t\ge1,\;r\neq1
\end{equation}

When lifespan $(T_{r}^{\max})$ of one vehicle is reached, it is removed from service, which vehicle-scrapping logic is defined by constraint (\ref{eq:VehScrap}). More specifically, if $t\ge T_{r}^{\max}$, vehicles scrapped at period $t$ correspond exactly to those purchased at $t-T_{r}^{\max}$, ensuring retirement at the end of their lifespan:
\begin{equation}\label{eq:VehScrap}
  N\!V_{rt}^{\mathrm{scrap}}=
  \begin{cases}
    N\!V_{r,t-T_{r}^{\max}}^{\mathrm{new}}, & t\ge T_{r}^{\max},\\[4pt]
    0, & t<T_{r}^{\max},
  \end{cases}
  \quad \forall r\neq1
\end{equation}

\subsubsection{Environmental Constraint}

The total carbon emissions from transportation, calculated as the product of the emission factor, distance, and fuel economy, must not exceed the maximum allowable limit $E\!M_{jt}^{\mathrm{ceil}}$ set by local government for each demand node $j$. Constraint (\ref{eq:Env}) ensures that environmental regulations are adhered to across the network:
\begin{equation}\label{eq:Env}
  \sum_{i,r}
  e m_{r}\,L_{ij}\,
  \frac{N\!V_{rt}}{F\!E_{r}}
  \;\le\; E\!M_{jt}^{\mathrm{ceil}},
  \quad \forall j,t
\end{equation}

\subsection{Enhancements to the Model Formulation with Hydrogen Hub Integration}
\label{subsec:hub}

The integration of hydrogen hubs restructures the transportation model by introducing an intermediary stage between production plants and demand points. Hub locations were identified using K‐means clustering based on geographic proximity and hydrogen demand distribution. This new hub‐based structure requires adjustments to variables, the objective function, and transportation constraints, explicitly separating supply-to-hub and hub-to-demand transportation, ensuring consistent alignment with hub assignments. For clarity and brevity, detailed mathematical formulations and constraints are provided in Appendix~A. The key modifications summarized here include:

\begin{enumerate}[label=\arabic*)]

\item \textbf{New Coefficients Introduced}

$k\in{K}$ is added to represent the hydrogen hub locations, added as an intermediary between supply points and demand points.  

$c_j$ is introduced to represent clustering-based parameters, assigned to determine hub coverage area, who assigns each demand node $j$ to a unique \textit{hu k b}, ensuring that all hydrogen delivered to $j$ must pass through its designated hub, in other words, if $c_j=k$, then demand node $j$ belongs to the cluster associated with hub $k$, enforcing a structured hub-based allocation.

\item \textbf{Parameters and Variables Adjustments}

The original transportation flow variable $V_{ijrt}$ is split into $V_{ijrt}$ for hydrogen flow from supply points to hubs, and $V_{kjrt}$ for hydrogen flow from hubs to demand points.  

The binary pipeline variables $B\!H_{ijt}$ and $B\!H_{ijt}^{\mathrm{new}}$ is updated accordingly, where $B\!H_{ikt}$ represents pipeline existing between supply points and hubs, and $B\!H_{kjt}$ represents pipeline existing between hubs and demand points; $B\!H_{ikt}^{\mathrm{new}}$ represents pipeline is built between supply points and hubs, and $B\!H_{kjt}^{\mathrm{new}}$ represents pipeline is built between hubs and demand points.  

The cost term $C_{ijt}^{KH}$ is split to reflect the two-stage transportation structure, where $C_{ikt}^{KH}$ represents the cost of pipeline construction from supply points to hubs, and $C_{kjt}^{KH}$ represents the cost of pipeline construction from hubs to demand points.

\item \textbf{Objective Function Modifications}

To prevent redundant calculations, the total hydrogen delivered is now represented as the summation of $V_{kjrt}$ showed in Equation~\eqref{eqA2}, ensuring that only the final hydrogen deliveries from hubs to demand nodes are counted.  

The Hydrogen Pipeline Investment Cost is modified in Equation~\eqref{eqA5} to reflect the introduction of hubs, splitting the original direct supply-to-demand pipeline investment into two stages. This adjustment ensures that the cost model accurately captures the two-stage transportation structure introduced in Mode~2.  

The Fuel Cost and Driver Labor Cost are modified in Equation~\eqref{eqA8} and \eqref{eqA9} to exclusively account for vehicle-based transport from hubs to demand points. This change is necessary because, in Mode~2, the first transportation stage, $i$ to $k$, is strictly pipeline-based, meaning vehicles are only used in the second stage, $k$ to $j$.  

The Pipeline Maintenance Cost and Hydrogen Loss Cost are restructured into separate cost components in Equation~\eqref{eqA10} and \eqref{eqA12}. These updates ensure a clearer distinction between pipeline and vehicle-based transportation costs, which reflects the two-stage transportation structure.

\item \textbf{New Constraints Introduced}

The Hub Balance Constraint \eqref{eqA16} is added to ensure that the total inflow from supply nodes into any hub $k$ (via pipeline) must equal the total outflow from this hub to demand nodes.  

The Cluster Assignment Constraint \eqref{eqA21} is introduced to prevent multiple hubs from supplying the same demand node and to ensure strict adherence to the clustering assignment. Specifically, if $k\neq c_j$, then $V_{kjrt}=0$ and $B\!H_{kjt}=0$, meaning that demand node $j$ cannot receive hydrogen from any hub other than its assigned one, $c_j$, and no pipeline is allowed to exist along this route.

\item \textbf{Original Transportation Constraints Restructured}

Constraints \eqref{eqA22}–\eqref{eqA30} are restructured into a two-step process, where $i$ to $k$ represents hydrogen transport from supply points to hubs, restricted to pipelines, and $k$ to $j$ represents transport from hubs to demand points using either pipelines or vehicles. This restructuring explicitly separates pipeline-only and mixed-mode transport, ensuring that transportation planning and route allocation are properly enforced at each stage.
\end{enumerate}

\section{Hydrogen Transportation Scenarios in Texas}\label{sec:Texas}

The case study focuses on optimizing the hydrogen transportation in Texas by integrating real‐world data with projected demand trends. It begins by establishing supply and demand nodes based on Gulf Coast production facilities and anticipated hydrogen consumption in key metropolitan areas. To explore different transportation strategies, five scenarios are developed, incorporating variations in geographic distribution and network configuration. The analysis then evaluates the impact of pipeline deployment, demand dynamics, and hub integration on transportation planning and infrastructure development.

\subsection{Data Description}\label{ssec:data}

\subsubsection{Basic Data Collection}\label{sssec:basic}

To accurately model transportation efficiency and costs, parameters were sourced from various literature and official datasets. Where direct data was unavailable, reasonable assumptions were derived based on industry trends and comparable studies. Table~\ref{tab:CostParams} provides an overview of the assigned values for cost‐related parameters. The capital costs of vehicles ($C_{r0}^{KV}$), hydrogen pipeline construction cost ($C_{0}^{KH}$), and the discount rate ($\beta$) were directly adopted from \citet{Teichmann2012}, \citet{Welder2018}, and \citet{Bodal2020}, respectively, which studies provide established values for hydrogen transportation infrastructure. Other parameters, including fuel economy, maximum travel distance, and truck speed, were modified based on relevant literature. Specifically, the average truck speed ($S\!P_{r}$) was adjusted according to data from \citet{FHWA2019}, which provides freight speed estimates across different road types in the United States. Vehicle operational characteristics, including fuel efficiency ($F\!E_{r}$), allowable travel time ($A\!T_{r}$), and infrastructure lifespan ($T_{r}^{\max}$), were refined using insights from \citet{Teichmann2012}. Furthermore, load capacity ($V_{r}^{\mathrm{cap}}$) and loading/unloading time of each vehicle transportation method ($L\!T_{r}$) were modified based on data from \citet{Rong2024}. The fuel price ($P_{r0}^{\mathrm{fuel}}$) was established according to Texas fuel pricing data provided by the \citet{EIA2024}. The driver hourly wage ($P_{r0}^{\mathrm{wage}}$) was slightly adjusted based on wage information for heavy‐duty truck drivers in Texas from the \citet{BLS2023}. Collectively, these parameters—both directly adopted from established literature and refined through reasonable adjustments—ensure that when the mathematical model is executed, it accurately captures real‐world transportation operations without compromising computational efficiency.

\begin{table}[h!]
\centering
\caption{Cost‐Related Parameters}
\label{tab:CostParams}
\scriptsize
\renewcommand{\arraystretch}{1.05}
\begin{tabular}{lcccc}
\toprule
 & \textbf{Tube Trailer} & \textbf{Liquid Truck} & \textbf{LOHC‐Trailer} & \textbf{Pipeline}\\
\midrule
$C_{r0}^{KV}$ (\$/vehicle)    & \$271420 & \$173709 & \$86854  &            \\[2pt]
$C_{0}^{KH}$ (\$/km)          &          &          &          & \$1735904  \\[2pt]
$T_{r}^{\max}$ (year)         & 12       & 8        & 12       & 40         \\[2pt]
$A\!T_{r}$ (hour/day)         & 10       & 10       & 10       &            \\[2pt]
$F\!E_{r}$ (km/L)             & 2.86     & 2.86     & 2.86     &            \\[2pt]
$S\!P_{r}$ (km/hour)          & 80       & 80       & 80       &            \\[2pt]
$V_{r}^{\mathrm{cap}}$ (kg)   & 500      & 3500     & 1500     &            \\[2pt]
$L\!T_{r}$ (hour)             & 2        & 3        & 2        &            \\[2pt]
$P_{r0}^{\mathrm{fuel}}$ (\$/L)& 0.71    & 0.71     & 0.71     &            \\[2pt]
$P_{r0}^{\mathrm{wage}}$ (\$/hour)& 28   & 26       & 28       &            \\[2pt]
$\beta$ (discount rate)       & 6.6\%    & 6.6\%    & 6.6\%    &            \\
\bottomrule
\end{tabular}
\end{table}

\subsubsection{Supply and Demand Forecast}\label{sssec:forecast}

To define supply and demand locations, the model applies straight‐line distances between two points. The coordinate centers of the hydrogen supply county and the hydrogen demand county are designated as supply nodes and demand nodes, respectively. This approach provides a simplified but structured representation of the hydrogen distribution network, ensuring consistency in demand forecasting and transport modeling.

\begin{figure}[h!]
    \centering
     \includegraphics[scale=1]{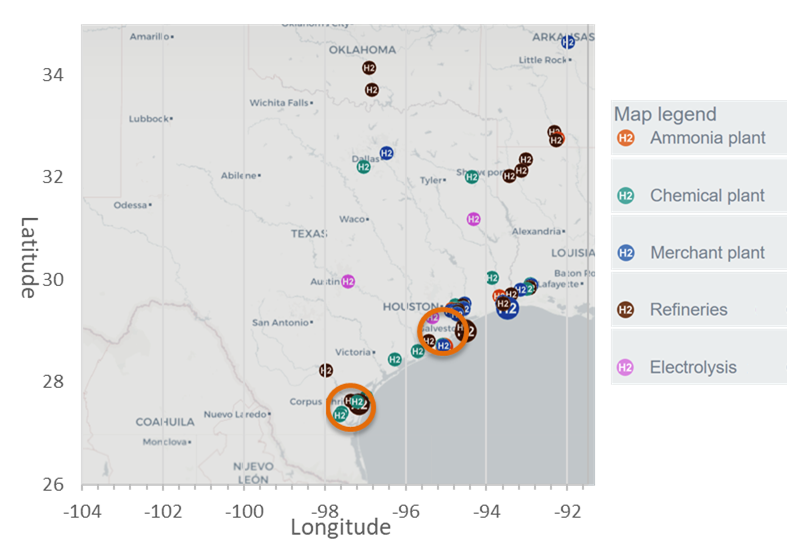}  
     \caption{\textit{Distribution of Hydrogen Production Plants in Texas}}
    \label{figure2}
\end{figure}

\noindent\textbf{Supply Points:}\\
A key observation is identified by analyzing the spatial distribution of existing and potential hydrogen production plants in Texas, categorized by different hydrogen production methods, as reported by the \citet{I3_2024}. Figure~\ref{figure2} illustrates this distribution, with hydrogen production facilities represented by colored dots, where each color corresponds to a specific production method. The majority of these facilities are concentrated along the Gulf Coast, forming a key hydrogen supply corridor. Building on this insight, our study selects Harris County (home to Houston) and Nueces County (home to Corpus Christi) as supply points, as these counties exhibit the highest density of hydrogen production facilities. In Figure~\ref{figure2}, these supply locations are highlighted with orange circles, specifically marking the large‐scale hydrogen production plants that serve as major contributors to regional supply. This selection ensures that the study accurately reflects real‐world hydrogen availability and infrastructure potential in Texas.

\noindent\textbf{Supply Prediction:}\\
To mitigate optimization challenges arising from potential variations in production output and supply shortages, the study assumes that the total hydrogen supply in each period slightly exceeds the total demand. This assumption ensures sufficient supply capacity, thereby maintaining computational stability throughout the optimization process.

The hydrogen supply capacities at selected locations were estimated based on the density of hydrogen production plants, as illustrated in Figure~\ref{figure2}. Harris County, having approximately six hydrogen production plants, is projected to contribute 60\% of the total supply, while Nueces County, with approximately four plants, accounts for the remaining 40\%. These proportions reflect the relative concentration of production facilities and form the basis for the estimated supply allocations shown in Table~\ref{tab:SupplyAlloc}.

\begin{table}[h!]
\centering
\caption{Prediction of Supply Capacity Proportion over Supply Points}
\label{tab:SupplyAlloc}
\scriptsize
\renewcommand{\arraystretch}{1.05}
\begin{tabular}{lcc}
\toprule
\textbf{Supply Points (Counties)} & \textbf{Harris} & \textbf{Nueces}\\
\midrule
Approximate number of hydrogen production plants & 6  & 4  \\[2pt]
Supply capacity proportion forecast              & 60\% & 40\% \\
\bottomrule
\end{tabular}
\end{table}

\begin{figure}[h!]
    \centering
     \includegraphics[scale=1]{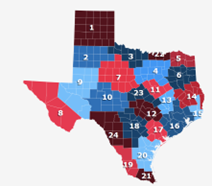}  
     \caption{\textit{Texas Regional Councils}}
    \label{figure3}
\end{figure}

\noindent\textbf{Demand Points:}\\
To Start with, our study divides Texas into 24 regions according to the Texas Association of Regional Councils \citep{TARC2025}, with each region comprising several counties, as shown in Figure~\ref{figure3}. Subsequently, different groups of demand points were selected for each scenario.

\begin{enumerate}[label=\arabic*)]
\item \textbf{For Scenarios 1,4 \&5:}\\
Three regions—Region 4, Region 12, and Region 18—were chosen, covering major metropolitan areas such as Dallas, Austin, and San Antonio. Harris County, home to Houston, was omitted as it was already designated as a supply point. Within each selected region, the four counties with the highest populations were identified based on Texas population data \citep{TexasDataLab2024}, resulting in a total of 12 demand points.

\item \textbf{For Scenarios 2 \&3:}\\
To introduce greater variation in distance and provide contrast, three regions located relatively far apart—Region 1, Region 3, and Region 8—were selected. Following the same criteria, the counties with the largest populations within these regions were identified, forming another set of 12 demand points.
\end{enumerate}

\noindent\textbf{Demand Prediction:}\\
The hydrogen demand for each county in Texas is projected annually from 2025 to 2050, based on several key factors. The county population is estimated using Texas's projected growth rate, starting from the 2024 population data \citep{TexasDataLab2024}. Average yearly mileage follows the Federal Highway Administration's estimate, where Americans drive an average of 22,954 km per year. A commuter ratio of 45\% is assumed, representing the proportion of the population expected to use hydrogen fuel cell vehicles for commuting purposes \citep{Ogumerem2018}. The FCEV efficiency is based on a standardized consumption rate of 1 kg of hydrogen per 100 km \citep{DiPierro2024}. The FCEV sales share is projected based on trends in the zero-emission vehicle (ZEV) sales share, and the FCEV share within ZEVs in California over the next 20 years \citep{Vijayakumar2022}, with adjustments to reflect Texas’s significantly lower current EV adoption rate, which is near zero \citep{ICCT2023}. Table~\ref{tab:FECVShare} presents the estimated FCEV sales share rates for Texas from 2025 to 2050.

\begin{table}[h!]
\centering
\caption{Texas FECV Sales Share Forecast (2030–2050)}
\label{tab:FECVShare}
\scriptsize
\renewcommand{\arraystretch}{1.05}
\begin{tabular}{lcccc}
\toprule
\textbf{Year} & \textbf{ZEV Sale/ALL (\%)} & \textbf{FECV/ZEV (\%)} & \textbf{FECV/ALL (\%) = $k$}\\
\midrule
2025 & 5   & 0.1 & 0.005\\
2030 & 24  & 10  & 2.4  \\
2035 & 43  & 20  & 8.6  \\
2040 & 62  & 30  & 18.6 \\
2045 & 81  & 40  & 32.4 \\
2050 & 100 & 50  & 50   \\
\bottomrule
\end{tabular}
\end{table}

Utilizing these inputs, the hydrogen demand for FCEVs in each county per year is determined by factoring in population size, commuter ratio, FCEV sales share rate, average mileage, and vehicle efficiency. This approach, guided by \citet{Ogumerem2018}, integrates demographic trends and vehicle fuel consumption to estimate hydrogen demand:
\begin{equation}
\resizebox{\textwidth}{!}{$
\text{Hydrogen Demand (kg)}
  =\text{County Population}\times
   \text{Commuter Ratio}\times
   \text{FCEV Sales Share}\times
   \text{Average Yearly Mileage}\times
   \text{FCEV Efficiency}
$}
\label{eq:demand_general}
\end{equation}

Simplified further with the assumed values, where $k$ represents the FCEV sales share for the corresponding period:
\begin{equation}
\text{Hydrogen Demand (kg)}
  = \text{County Population}\times k \times 103.293\,
   \frac{\text{kg}}{\mathit{year}_{\text{Hydrogen}}}
\label{eq:demand_simple}
\end{equation}

\noindent\textbf{Identification of Hydrogen Hub Locations for Scenario~5:}\\
The locations of hydrogen hubs implemented in Scenario~5 are determined by applying K‐means clustering to the 12 demand points defined for Scenarios~1,4~\&5. The clustering approach grouped these points based on geographic proximity and similarity in hydrogen demand levels, resulting in three optimal hub locations tailored specifically for Scenario~5. This ensures each hub effectively represents and serves its assigned cluster of demand nodes.

\subsubsection{Scenarios}\label{sssec:scenarios}

The five scenarios are designed to examine how demand distribution, supply and demand levels, pipeline construction schedules, and transportation mode selection influence hydrogen transportation efficiency and infrastructure planning. These variations allow the model to assess different network configurations, balancing economic feasibility, and resource allocation under varying conditions. Table~\ref{tab:ScenarioCmp} summarizes the key differences among these scenarios.

In terms of demand point distribution, \textbf{S1} and \textbf{S4} use proximal demand points, while \textbf{S2} and \textbf{S3} use distant demand points. The difference between \textbf{S2} and \textbf{S3} lies in supply–demand quantities—\textbf{S2} retains the same values as \textbf{S1}, whereas \textbf{S3} adjusts them to reflect different consumption levels. Regarding pipeline construction periods, \textbf{S1}, \textbf{S2}, \textbf{S3}, and \textbf{S5} assume a 1‐year construction period, while \textbf{S4} extends it to 2 years, affecting infrastructure deployment timelines and cost distribution. For transportation modes, \textbf{S1}, \textbf{S2}, \textbf{S3}, and \textbf{S4} operate under Mode 1, where hydrogen is transported directly from supply points to demand points; \textbf{S5}, as Mode 2, includes hydrogen hubs under the two‐stage distribution system.

\begin{table}[h!]
\centering
\caption{Similarities and Differences of the Five Scenarios}
\label{tab:ScenarioCmp}
\scriptsize
\renewcommand{\arraystretch}{1.05}   
\begin{tabularx}{\linewidth}{l X X X X}
\toprule
\textbf{Scenario} &
\textbf{Straight Distance from Supply to Demand Points} &
\textbf{Demand \& Supply Level} &
\textbf{Construction Period for One Pipeline} &
\textbf{Hydrogen Hub}\\
\midrule
S1 & Proximal & High & 1~year & No Hub\\
S2 & Distant  & Low  & 1~year & No Hub\\
S3 & Distant  & High & 1~year & No Hub\\
S4 & Proximal & High & 2~years & No Hub\\
S5 & Proximal & High & 1~year & With Hub\\
\bottomrule
\end{tabularx}
\end{table}

\subsection{Results and Discussion}\label{sec:results}

Results are obtained by running the model using the \textsc{Gurobi} optimizer on a system equipped with an \textit{Intel Core i9-13980HX} processor (24 physical cores, 32 threads) and \textit{Windows 11}. The implementation utilizes key computational packages, including \texttt{NumPy}, \texttt{Pandas}, \texttt{SciPy}, \texttt{Geopy}, \texttt{Matplotlib}, and \texttt{GurobiPy}. The model achieves an optimal solution in \textbf{78.08 s} (Scenario 1), demonstrating its efficiency in handling large-scale hydrogen transportation optimization problems.

The result analysis first examines the temporal evolution of hydrogen transportation method reliance, which highlights the increasing role of pipeline transport as infrastructure expands. Next, demand-level analysis explores how variations in hydrogen consumption impact transport method selection. Following this, a distance-level analysis assesses the influence of supply-demand distance on transportation strategies, illustrating the trade-offs between vehicle reliance and pipeline deployment. The study then evaluates pipeline construction period effects, investigating how different infrastructure deployment schedules affect network efficiency. Finally, the impact of hydrogen hub integration is analyzed, comparing direct transport networks with hub-based distribution to assess their respective benefits in cost and flexibility.

\subsubsection{Performance Comparison Among Scenarios S1, S2, S3, and S4 (Mode 1)}\label{subsec:S1S4}
This section compares Scenarios~S1, S2, S3, and S4, all under \textbf{Mode 1}, focusing on the impact of distance, demand levels, and pipeline construction time on transportation costs and system efficiency.

\begin{figure}[h!]
    \centering
     \includegraphics[scale=0.7]{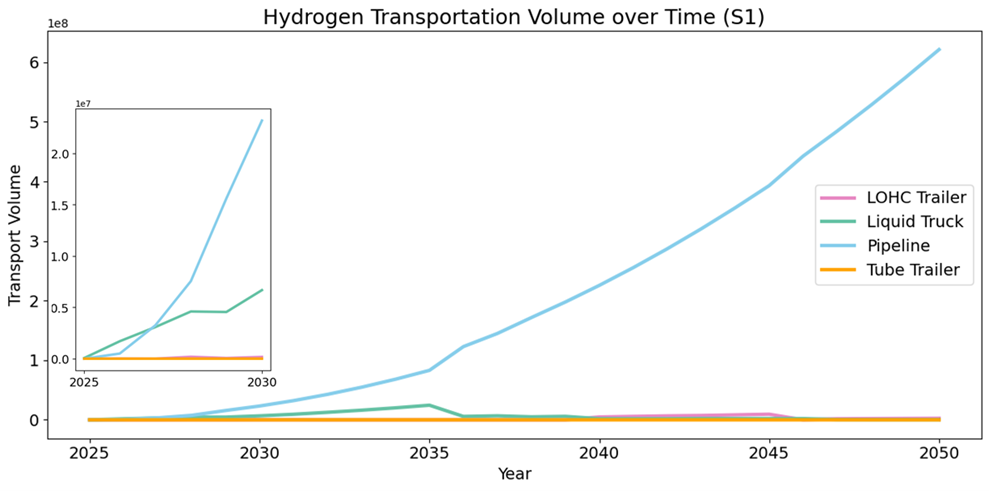}  
     \caption{\textit{Hydrogen transportation volume over time periods}}
    \label{figure4}
\end{figure}

\noindent\textbf{Temporal Trend Analysis}\\
Over time, pipeline transport becomes the dominant mode for hydrogen delivery, gradually replacing vehicle-based transport. Figure~\ref{figure4} illustrates this trend under Scenario 1, where the pipeline construction period is set to one year. Initially, multiple transportation modes, including liquid trucks and tube trailers, contribute to hydrogen distribution. However, as more pipelines are constructed, the transport volume via pipelines increases significantly, while reliance on vehicles declines.

This shift is evident as pipeline transport sees exponential growth after 2030, corresponding to rising hydrogen demand. In contrast, vehicle-based transport gradually declines and contributes negligibly in later years. The transition indicates that, once pipelines are established, they handle most of the hydrogen transport, reducing the need for alternative methods. The model captures this progression by optimizing transport allocation over time, leading to a near-total reliance on pipelines by 2050.

\begin{figure}[h!]
    \centering
     \includegraphics[scale=0.9]{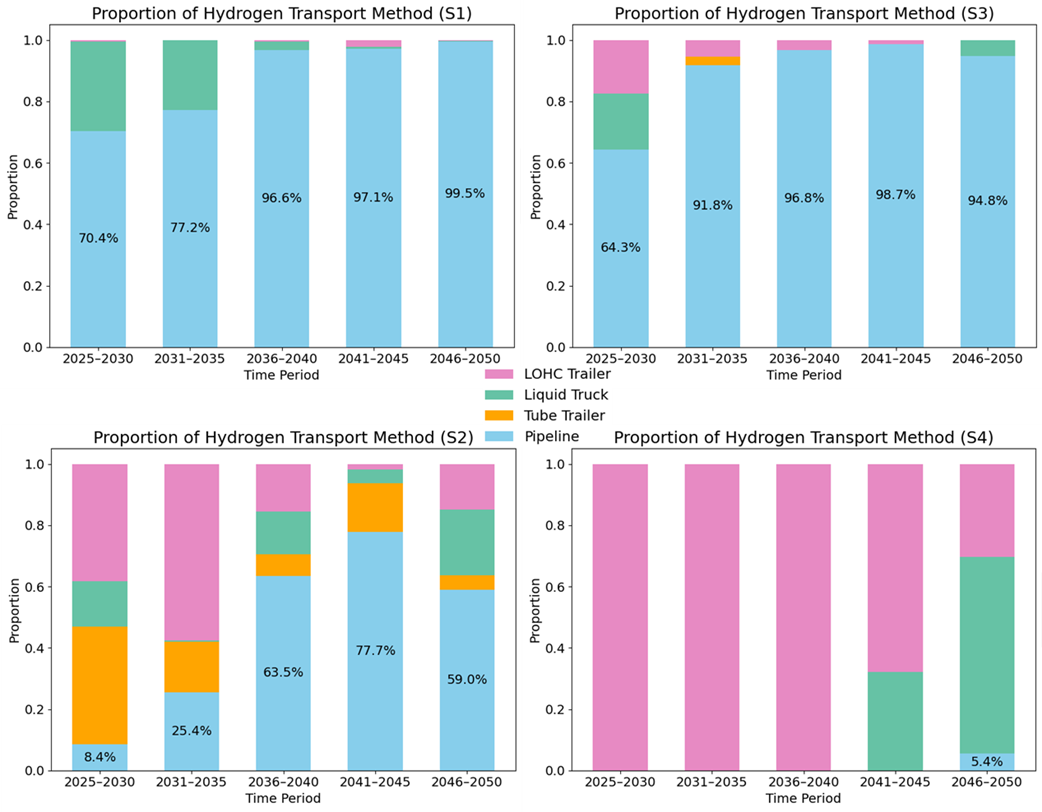}  
     \caption{\textit{ Hydrogen transportation proportion over time periods S1, S2, S3, S4}}
    \label{figure5}
\end{figure}

\noindent\textbf{Demand Level Analysis}\\
As hydrogen demand increases, reliance on pipeline transportation becomes more dominant. This trend is evident when comparing \textbf{S2} and \textbf{S3}, where \textbf{S3} with higher demand levels exhibits a significantly greater pipeline share across all time periods. As shown in Figure~\ref{figure5}, which presents the proportion of hydrogen transported by different methods over time, pipeline transport in \textbf{S3} accounts for 64.3\% of the total hydrogen transported in the 2025–2030 period, whereas \textbf{S2} relies on pipelines for only 8.4\%, primarily depending on vehicle-based transportation. Over time, pipeline dependence in \textbf{S3} rises to 94.8\% by 2046–2050, while \textbf{S2} remains at a lower 59.0\%, still maintaining a notable share of liquid and tube trailer transport.

This discrepancy can be attributed to economies of scale, where higher demand justifies pipeline construction, ultimately resulting in a 23\% reduction in optimal levelized costs in \textbf{S3} compared to \textbf{S2} with lower demand that relies more on vehicle transportation.

\begin{figure}[h!]
    \centering
     \includegraphics[scale=1]{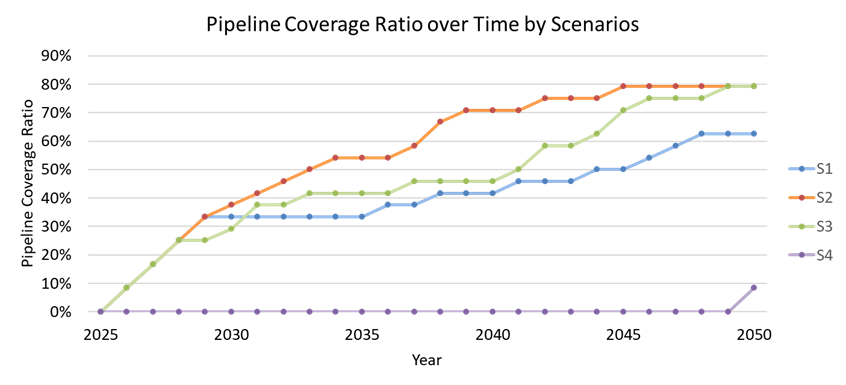}  
     \caption{\textit{Pipeline Coverage Ratio over Time}}
    \label{figure6}
\end{figure}

\noindent\textbf{Distance Level Analysis}\\
As shown in Figure~\ref{figure5}, Scenario~3 (\textbf{S3}), with longer distances between supply and demand points, initially has a lower share of hydrogen transported via pipelines compared to Scenario~1 (\textbf{S1}). However, after 2030, the trend shifts as \textbf{S3} experiences a more rapid increase in pipeline adoption. This suggests that while shorter distances allow for a higher initial pipeline utilization, long‐distance scenarios progressively rely more on pipelines over time. Figure~\ref{figure6} further supports this observation, demonstrating that pipeline coverage in \textbf{S3} eventually surpasses \textbf{S1}. The pipeline coverage ratio is defined as the number of pipelines present in each period divided by the total number of pipelines required for complete coverage.Where the complete coverage in this Texas study is $12\ {demand}\ {nodes}\times 2\ {spply}\ {nodes}=24$, representing all supply points connected to all demand points through pipelines. By the final period, pipeline coverage in the long‐distance scenario is 27\% higher than in the short‐distance counterpart. This indicates that extended transportation distances amplify the benefits of pipeline infrastructure, leading to greater pipeline deployment in later stages.

\noindent\textbf{Pipeline Construction Periods Analysis}\\
The construction period significantly impacts pipeline deployment, with \textbf{S4}, having a two-year construction period, displaying much lower pipeline coverage than \textbf{S1}, \textbf{S2}, and \textbf{S3}, which all follow a one-year period (Figure~\ref{figure6}). While the other scenarios steadily expand their pipeline networks, \textbf{S4} remains without pipelines for almost the entire timeline and begins construction only in the final stages. This demonstrates that longer construction periods delay pipeline implementation, leading to substantially lower coverage compared to scenarios with shorter timelines.

\subsubsection{Evaluating the Impact of Hydrogen Hub Integration: S1 (Mode 1) vs S5 (Mode 2)}
\label{subsec:hubImpact}

This section compares Scenario \textbf{S1} (Mode 1) with Scenario \textbf{S5} (Mode 2), evaluating the effect of hydrogen hub integration on transportation efficiency and cost.

\begin{figure}[h!]
    \centering
     \includegraphics[scale=0.9]{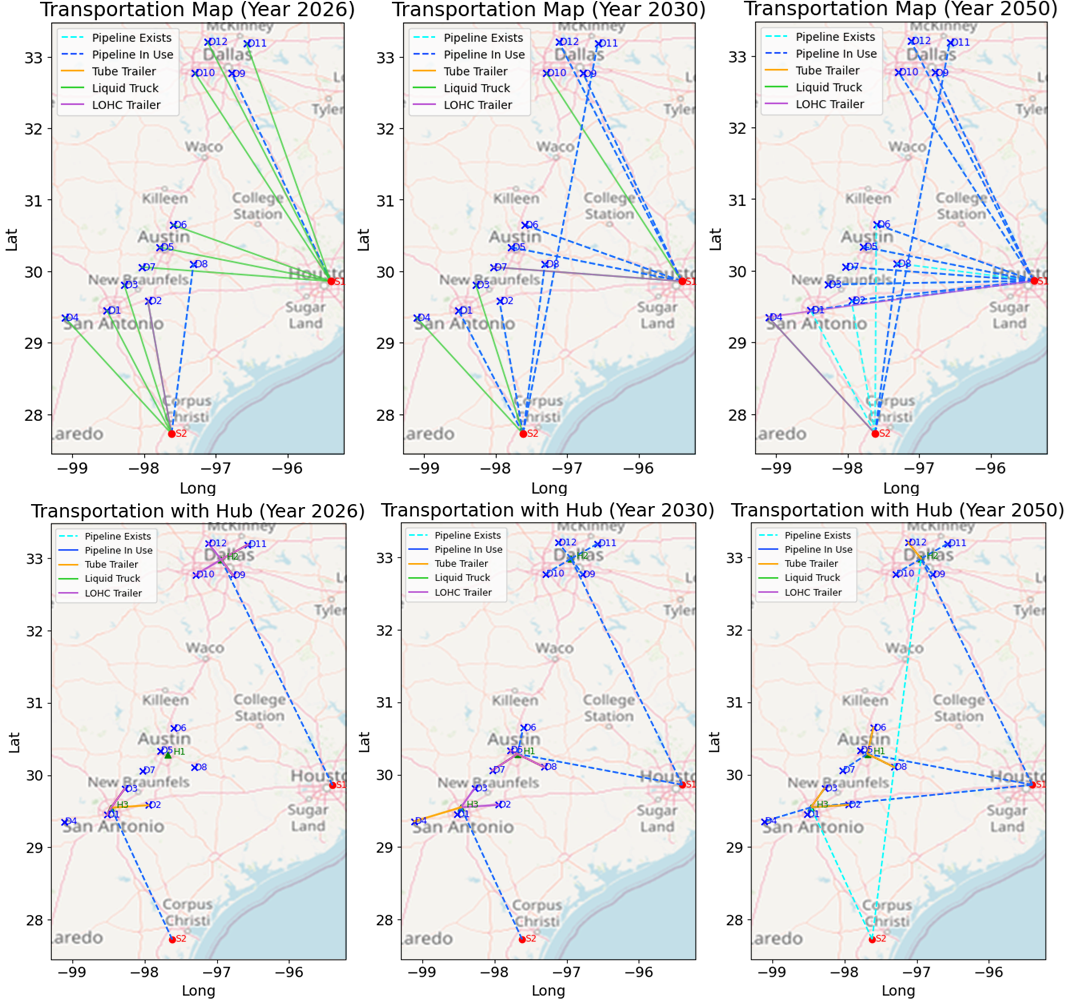}  
     \caption{\textit{Development of Hydrogen Transportation Networks Over Time: Direct Transport (S1) vs Hub-Based Transport (S5)}}
    \label{figure7}
\end{figure}

The transportation networks in \textbf{S1} and \textbf{S5} follow distinct structural developments over time, as shown in Figure~\ref{figure7}. In \textbf{S1}, the network transitions from an initial mix of pipelines and vehicle transport to a predominantly pipeline-based system after 2030, simplifying into a direct, point-to-point structure between supply and demand nodes. In contrast, \textbf{S5} exhibits a more intricate and interconnected topology due to the integration of hydrogen hubs. These hubs facilitate a multi-stage distribution process, maintaining vehicle-based transport for longer periods and enabling greater flexibility in route adaptation.

\begin{figure}[h!]
    \centering
     \includegraphics[scale=1]{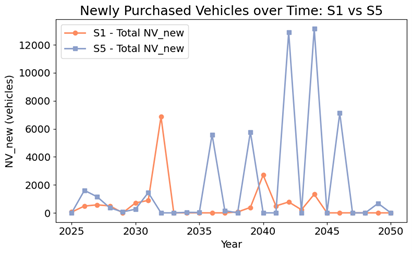}  
     \caption{\textit{Newly Purchased Vehicles over Time: S1 vs S5}}
    \label{figure8}
\end{figure}

Figure~\ref{figure8} illustrates how these structural differences influence vehicle deployment. In \textbf{S1}, the number of newly purchased vehicles peaks early but declines sharply after 2030, reflecting the increasing reliance on pipelines. By contrast, \textbf{S5} maintains recurring peaks in vehicle purchases throughout the planning horizon, with some late‐stage peaks exceeding twice the highest vehicle purchase level observed in \textbf{S1}. This sustained demand for vehicles in \textbf{S5} highlights the ongoing need for flexible transport solutions due to hub‐based redistribution, whereas \textbf{S1}’s direct pipeline approach gradually eliminates vehicle dependence.

These observations confirm that hydrogen hubs significantly alter transportation patterns by maintaining the need for vehicle transport even as pipelines expand. In \textbf{S1}, vehicle usage diminishes as the network transitions to full pipeline reliance, whereas \textbf{S5} retains vehicle transport as an integral component due to the hub structure. This difference suggests that Mode 2 allows for a more distributed and adaptable transportation network, where hydrogen can be redirected through hubs as needed, rather than relying solely on rigid direct pipeline connections.


\section{Conclusion}

This study provides a comprehensive assessment of how infrastructure phasing, transport mode selection, and spatial network design jointly influence the long-term performance of hydrogen distribution systems. Through a series of scenario based simulations, we uncover several critical insights. First, infrastructure sequencing plays a decisive role: even modest delays in pipeline construction, such as extending lead times from one to two years, can reduce network coverage by over 60\% and significantly elevate reliance on less efficient vehicle based transport. Second, both the scale and geographic dispersion of hydrogen demand shape the optimal transport strategy. Pipelines become increasingly dominant in high-volume, long-distance scenarios and offer substantial cost savings, while vehicle transport remains essential in early deployment stages or under constrained infrastructure timelines. Third, the integration of hydrogen hubs adds strategic flexibility by enabling a decentralized, two stage logistics structure that supports vehicle use for last mile delivery and facilitates redistribution across regions with variable demand. Collectively, these findings underscore the importance of coordinated, forward looking infrastructure planning that aligns investment timing, asset lifecycles, and multimodal logistics to meet evolving demand patterns. While this work is grounded in the Texas context, the modeling framework is generalizable and can be adapted for other regions facing similar challenges in hydrogen supply chain planning. The insights generated here offer a practical foundation for designing scalable, resilient, and economically viable hydrogen transportation systems that can support broader energy transition objectives.

\bibliographystyle{elsarticle-harv} 
\bibliography{references}

\begin{thebibliography}{37}
\expandafter\ifx\csname natexlab\endcsname\relax\def\natexlab#1{#1}\fi
\providecommand{\url}[1]{\texttt{#1}}
\providecommand{\href}[2]{#2}
\providecommand{\path}[1]{#1}
\providecommand{\DOIprefix}{doi:}
\providecommand{\ArXivprefix}{arXiv:}
\providecommand{\URLprefix}{URL: }
\providecommand{\Pubmedprefix}{pmid:}
\providecommand{\doi}[1]{\href{http://dx.doi.org/#1}{\path{#1}}}
\providecommand{\Pubmed}[1]{\href{pmid:#1}{\path{#1}}}
\providecommand{\bibinfo}[2]{#2}
\ifx\xfnm\relax \def\xfnm[#1]{\unskip,\space#1}\fi
\bibitem[{Arnaboldi et~al.(2014)Arnaboldi, Azzone and Giorgino}]{Arnaboldi2014}
\bibinfo{author}{Arnaboldi, M.}, \bibinfo{author}{Azzone, G.}, \bibinfo{author}{Giorgino, M.}, \bibinfo{year}{2014}.
\newblock \bibinfo{title}{Performance Measurement and Management for Engineers}.
\newblock \bibinfo{publisher}{Academic Press}.
\bibitem[{Bade et~al.(2024)Bade, Tomomewo, Meenakshisundaram, Ferron and Oni}]{Bade2024}
\bibinfo{author}{Bade, S.O.}, \bibinfo{author}{Tomomewo, O.S.}, \bibinfo{author}{Meenakshisundaram, A.}, \bibinfo{author}{Ferron, P.}, \bibinfo{author}{Oni, B.A.}, \bibinfo{year}{2024}.
\newblock \bibinfo{title}{Economic, social, and regulatory challenges of green hydrogen production and utilization in the us: A review}.
\newblock \bibinfo{journal}{International Journal of Hydrogen Energy} \bibinfo{volume}{49}, \bibinfo{pages}{314--335}.
\bibitem[{B{\o}dal et~al.(2020)B{\o}dal, Mallapragada, Botterud and Korp{\aa}s}]{Bodal2020}
\bibinfo{author}{B{\o}dal, E.F.}, \bibinfo{author}{Mallapragada, D.}, \bibinfo{author}{Botterud, A.}, \bibinfo{author}{Korp{\aa}s, M.}, \bibinfo{year}{2020}.
\newblock \bibinfo{title}{Decarbonization synergies from joint planning of electricity and hydrogen production: A texas case study}.
\newblock \bibinfo{journal}{International Journal of Hydrogen Energy} \bibinfo{volume}{45}, \bibinfo{pages}{32899--32915}.
\bibitem[{{Bureau of Labor Statistics}(2023)}]{BLS2023}
\bibinfo{author}{{Bureau of Labor Statistics}}, \bibinfo{year}{2023}.
\newblock \bibinfo{title}{Occupational employment and wage statistics}.
\newblock \URLprefix \url{https://www.bls.gov/oes/2023/may/oes533032.htm}.
\bibitem[{Chen et~al.(2021)Chen, Gu, Tang, Wang and Wu}]{Chen2021}
\bibinfo{author}{Chen, Q.}, \bibinfo{author}{Gu, Y.}, \bibinfo{author}{Tang, Z.}, \bibinfo{author}{Wang, D.}, \bibinfo{author}{Wu, Q.}, \bibinfo{year}{2021}.
\newblock \bibinfo{title}{Optimal design and techno-economic assessment of low-carbon hydrogen supply pathways for a refueling station located in shanghai}.
\newblock \bibinfo{journal}{Energy} \bibinfo{volume}{237}, \bibinfo{pages}{121584}.
\bibitem[{Dautel et~al.(2024)Dautel, Thakur and Elberry}]{Dautel2024}
\bibinfo{author}{Dautel, J.L.}, \bibinfo{author}{Thakur, J.}, \bibinfo{author}{Elberry, A.M.}, \bibinfo{year}{2024}.
\newblock \bibinfo{title}{Enabling industrial decarbonization: A milp optimization model for low-carbon hydrogen supply chains}.
\newblock \bibinfo{journal}{International Journal of Hydrogen Energy} \bibinfo{volume}{77}, \bibinfo{pages}{863--891}.
\bibitem[{Deng et~al.(2023)Deng, Bian, Zhou and Ding}]{Deng2023}
\bibinfo{author}{Deng, M.}, \bibinfo{author}{Bian, B.}, \bibinfo{author}{Zhou, Y.}, \bibinfo{author}{Ding, J.}, \bibinfo{year}{2023}.
\newblock \bibinfo{title}{Distributionally robust production and replenishment problem for hydrogen supply chains}.
\newblock \bibinfo{journal}{Transportation Research Part E: Logistics and Transportation Review} \bibinfo{volume}{179}, \bibinfo{pages}{103293}.
\bibitem[{Di~Pierro et~al.(2024)Di~Pierro, Bitsanis, Tansini, Bonato, Martini and Fontaras}]{DiPierro2024}
\bibinfo{author}{Di~Pierro, G.}, \bibinfo{author}{Bitsanis, E.}, \bibinfo{author}{Tansini, A.}, \bibinfo{author}{Bonato, C.}, \bibinfo{author}{Martini, G.}, \bibinfo{author}{Fontaras, G.}, \bibinfo{year}{2024}.
\newblock \bibinfo{title}{Fuel cell electric vehicle characterisation under laboratory and in‐use operation}.
\newblock \bibinfo{journal}{Energy Reports} \bibinfo{volume}{11}, \bibinfo{pages}{611--623}.
\bibitem[{Dogliani et~al.(2024)Dogliani, Canha, Elberry and Thakur}]{Dogliani2024}
\bibinfo{author}{Dogliani, P.}, \bibinfo{author}{Canha, A.N.R.R.}, \bibinfo{author}{Elberry, A.M.}, \bibinfo{author}{Thakur, J.}, \bibinfo{year}{2024}.
\newblock \bibinfo{title}{Multi-option analytical modeling of levelized costs across various hydrogen supply chain nodes}.
\newblock \bibinfo{journal}{International Journal of Hydrogen Energy} \bibinfo{volume}{70}, \bibinfo{pages}{737--755}.
\bibitem[{Feng et~al.(2024)Feng, Manier and Manier}]{Feng2024}
\bibinfo{author}{Feng, L.}, \bibinfo{author}{Manier, H.}, \bibinfo{author}{Manier, M.A.}, \bibinfo{year}{2024}.
\newblock \bibinfo{title}{Designing a centralized storage hydrogen supply chain network with multi-period and bi-objective optimization}.
\newblock \bibinfo{journal}{Computers \& Chemical Engineering} \bibinfo{volume}{190}, \bibinfo{pages}{108820}.
\bibitem[{Forghani et~al.(2023)Forghani, Kia and Nejatbakhsh}]{Forghani2023}
\bibinfo{author}{Forghani, K.}, \bibinfo{author}{Kia, R.}, \bibinfo{author}{Nejatbakhsh, Y.}, \bibinfo{year}{2023}.
\newblock \bibinfo{title}{A multi-period sustainable hydrogen supply chain model considering pipeline routing and carbon emissions: The case study of oman}.
\newblock \bibinfo{journal}{Renewable and Sustainable Energy Reviews} \bibinfo{volume}{173}, \bibinfo{pages}{113051}.
\bibitem[{Saldanha-da Gama(2022)}]{Saldanha2022}
\bibinfo{author}{Saldanha-da Gama, F.}, \bibinfo{year}{2022}.
\newblock \bibinfo{title}{Facility location in logistics and transportation: An enduring relationship}.
\newblock \bibinfo{journal}{Transportation Research Part E: Logistics and Transportation Review} \bibinfo{volume}{166}, \bibinfo{pages}{102903}.
\bibitem[{Ganter et~al.(2024)Ganter, Gabrielli and Sansavini}]{Ganter2024}
\bibinfo{author}{Ganter, A.}, \bibinfo{author}{Gabrielli, P.}, \bibinfo{author}{Sansavini, G.}, \bibinfo{year}{2024}.
\newblock \bibinfo{title}{Near-term infrastructure rollout and investment strategies for net-zero hydrogen supply chains}.
\newblock \bibinfo{journal}{Renewable and Sustainable Energy Reviews} \bibinfo{volume}{194}, \bibinfo{pages}{114314}.
\bibitem[{Hermesmann et~al.(2023)Hermesmann, Tsiklios and M{\"u}ller}]{Hermesmann2023}
\bibinfo{author}{Hermesmann, M.}, \bibinfo{author}{Tsiklios, C.}, \bibinfo{author}{M{\"u}ller, T.E.}, \bibinfo{year}{2023}.
\newblock \bibinfo{title}{The environmental impact of renewable hydrogen supply chains: Local vs. remote production and long-distance hydrogen transport}.
\newblock \bibinfo{journal}{Applied Energy} \bibinfo{volume}{351}, \bibinfo{pages}{121920}.
\bibitem[{{International Council on Clean Transportation}(2023)}]{ICCT2023}
\bibinfo{author}{{International Council on Clean Transportation}}, \bibinfo{year}{2023}.
\newblock \bibinfo{title}{Electric vehicle market and policy developments in u.s. states, 2023}.
\newblock \URLprefix \url{https://theicct.org/publication/ev-ldv-us-major-markets-monitor-2023-june24/}. \bibinfo{note}{retrieved November 27, 2024}.
\bibitem[{{International Energy Agency}(2024)}]{IEA2024}
\bibinfo{author}{{International Energy Agency}}, \bibinfo{year}{2024}.
\newblock \bibinfo{title}{Global hydrogen review 2024}.
\newblock \URLprefix \url{https://www.iea.org/reports/global-hydrogen-review-2024}.
\bibitem[{{I\textsuperscript{3} Industrial Innovation Initiative}(2024)}]{I3_2024}
\bibinfo{author}{{I\textsuperscript{3} Industrial Innovation Initiative}}, \bibinfo{year}{2024}.
\newblock \bibinfo{title}{The landscape of clean hydrogen}.
\newblock \URLprefix \url{https://industrialinnovation.org/resources/the-landscape-of-clean-hydrogen/}.
\bibitem[{Kumar and Lim(2022)}]{KumarLim2022}
\bibinfo{author}{Kumar, S.S.}, \bibinfo{author}{Lim, H.}, \bibinfo{year}{2022}.
\newblock \bibinfo{title}{An overview of water electrolysis technologies for green hydrogen production}.
\newblock \bibinfo{journal}{Energy Reports} \bibinfo{volume}{8}, \bibinfo{pages}{13793--13813}.
\bibitem[{Leonard(2024)}]{Leonard2024}
\bibinfo{author}{Leonard, S.}, \bibinfo{year}{2024}.
\newblock \bibinfo{title}{Hydrogen fueling stations get funding for texas-california corridor}.
\newblock \URLprefix \url{https://www.truckingdive.com/news/hydrogen-fueling-stations-funding-texas-california-corridor/704291/}.
\bibitem[{Li et~al.(2021)Li, Al~Chami, Manier, Manier and Xue}]{Li2021}
\bibinfo{author}{Li, L.}, \bibinfo{author}{Al~Chami, Z.}, \bibinfo{author}{Manier, H.}, \bibinfo{author}{Manier, M.A.}, \bibinfo{author}{Xue, J.}, \bibinfo{year}{2021}.
\newblock \bibinfo{title}{Incorporating fuel delivery in network design for hydrogen fueling stations: Formulation and two metaheuristic approaches}.
\newblock \bibinfo{journal}{Transportation Research Part E: Logistics and Transportation Review} \bibinfo{volume}{152}, \bibinfo{pages}{102384}.
\bibitem[{Ogumerem et~al.(2018)Ogumerem, Kim, Kesisoglou, Diangelakis and Pistikopoulos}]{Ogumerem2018}
\bibinfo{author}{Ogumerem, G.S.}, \bibinfo{author}{Kim, C.}, \bibinfo{author}{Kesisoglou, I.}, \bibinfo{author}{Diangelakis, N.A.}, \bibinfo{author}{Pistikopoulos, E.N.}, \bibinfo{year}{2018}.
\newblock \bibinfo{title}{A multi-objective optimization for the design and operation of a hydrogen network for transportation fuel}.
\newblock \bibinfo{journal}{Chemical Engineering Research and Design} \bibinfo{volume}{131}, \bibinfo{pages}{279--292}.
\bibitem[{Perna et~al.(2023)Perna, Jannelli, Di~Micco, Romano and Minutillo}]{Perna2023}
\bibinfo{author}{Perna, A.}, \bibinfo{author}{Jannelli, E.}, \bibinfo{author}{Di~Micco, S.}, \bibinfo{author}{Romano, F.}, \bibinfo{author}{Minutillo, M.}, \bibinfo{year}{2023}.
\newblock \bibinfo{title}{Designing, sizing, and economic feasibility of a green hydrogen supply chain for maritime transportation}.
\newblock \bibinfo{journal}{Energy Conversion and Management} \bibinfo{volume}{278}, \bibinfo{pages}{116702}.
\bibitem[{{PwC}(2023)}]{PwC2023}
\bibinfo{author}{{PwC}}, \bibinfo{year}{2023}.
\newblock \bibinfo{title}{The green hydrogen economy - predicting the decarbonization agenda of tomorrow}.
\newblock \URLprefix \url{https://www.pwc.com/gx/en/industries/energy-utilities-resources/future-energy/green-hydrogen-cost.html}.
\bibitem[{Rong et~al.(2024)Rong, Chen, Li, Chen, Xie, Chen and Long}]{Rong2024}
\bibinfo{author}{Rong, Y.}, \bibinfo{author}{Chen, S.}, \bibinfo{author}{Li, C.}, \bibinfo{author}{Chen, X.}, \bibinfo{author}{Xie, L.}, \bibinfo{author}{Chen, J.}, \bibinfo{author}{Long, R.}, \bibinfo{year}{2024}.
\newblock \bibinfo{title}{Techno-economic analysis of hydrogen storage and transportation from hydrogen plant to terminal refueling station}.
\newblock \bibinfo{journal}{International Journal of Hydrogen Energy} \bibinfo{volume}{52}, \bibinfo{pages}{547--558}.
\bibitem[{Sizaire et~al.(2024)Sizaire, Lin and Gen{\c c}er}]{Sizaire2024}
\bibinfo{author}{Sizaire, P.}, \bibinfo{author}{Lin, B.}, \bibinfo{author}{Gen{\c c}er, E.}, \bibinfo{year}{2024}.
\newblock \bibinfo{title}{A novel hydrogen supply chain optimization model – case study of texas and louisiana}.
\newblock \bibinfo{journal}{International Journal of Hydrogen Energy} \bibinfo{volume}{78}, \bibinfo{pages}{401--420}.
\bibitem[{Teichmann et~al.(2012)Teichmann, Arlt and Wasserscheid}]{Teichmann2012}
\bibinfo{author}{Teichmann, D.}, \bibinfo{author}{Arlt, W.}, \bibinfo{author}{Wasserscheid, P.}, \bibinfo{year}{2012}.
\newblock \bibinfo{title}{Liquid organic hydrogen carriers as an efficient vector for the transport and storage of renewable energy}.
\newblock \bibinfo{journal}{International Journal of Hydrogen Energy} \bibinfo{volume}{37}, \bibinfo{pages}{18118--18132}.
\bibitem[{{Texas 2036 Data Lab}(2024)}]{TexasDataLab2024}
\bibinfo{author}{{Texas 2036 Data Lab}}, \bibinfo{year}{2024}.
\newblock \bibinfo{title}{Texas population data}.
\newblock \URLprefix \url{https://datalab.texas2036.org/user/download?id=328169}.
\bibitem[{{Texas Association of Regional Councils}(2025)}]{TARC2025}
\bibinfo{author}{{Texas Association of Regional Councils}}, \bibinfo{year}{2025}.
\newblock \bibinfo{title}{Regional councils}.
\newblock \URLprefix \url{https://txregionalcouncil.org/regional-councils/}.
\bibitem[{{U.S. Department of Energy}(2024)}]{DoEHub2024}
\bibinfo{author}{{U.S. Department of Energy}}, \bibinfo{year}{2024}.
\newblock \bibinfo{title}{Gulf coast hydrogen hub}.
\newblock \URLprefix \url{https://www.energy.gov/oced/gulf-coast-hydrogen-hub}. \bibinfo{note}{office of Clean Energy Demonstrations}.
\bibitem[{{U.S. Department of Transportation, Federal Highway Administration}(2019)}]{FHWA2019}
\bibinfo{author}{{U.S. Department of Transportation, Federal Highway Administration}}, \bibinfo{year}{2019}.
\newblock \bibinfo{title}{Table 3-11: Average truck speed by functional class and urban-rural designation}.
\newblock \URLprefix \url{https://ops.fhwa.dot.gov/freight/freight_analysis/nat_freight_stats/docs/11factsfigures/table3_11.htm}.
\bibitem[{{U.S. Energy Information Administration (EIA)}(2024)}]{EIA2024}
\bibinfo{author}{{U.S. Energy Information Administration (EIA)}}, \bibinfo{year}{2024}.
\newblock \bibinfo{title}{Texas all grades conventional retail gasoline prices}.
\newblock \URLprefix \url{https://www.eia.gov/dnav/pet/hist/LeafHandler.ashx?n=pet&s=emm_epm0u_pte_stx_dpg&f=m}.
\bibitem[{Vijayakumar(2022)}]{Vijayakumar2022}
\bibinfo{author}{Vijayakumar, V.}, \bibinfo{year}{2022}.
\newblock \bibinfo{title}{Understanding the evolution of hydrogen supply chains in the western united states: An optimization-based approach focusing on california as a future hydrogen hub}.
\newblock \URLprefix \url{https://theicct.org/publication/ev-ldv-us-major-markets-monitor-2023-june24/}. \bibinfo{note}{(Advisors: A. Jenn \& D. Sperling)}.
\bibitem[{Welder et~al.(2018)Welder, Ryberg, Kotzur, Grube, Robinius and Stolten}]{Welder2018}
\bibinfo{author}{Welder, L.}, \bibinfo{author}{Ryberg, D.S.}, \bibinfo{author}{Kotzur, L.}, \bibinfo{author}{Grube, T.}, \bibinfo{author}{Robinius, M.}, \bibinfo{author}{Stolten, D.}, \bibinfo{year}{2018}.
\newblock \bibinfo{title}{Spatio-temporal optimization of a future energy system for power-to-hydrogen applications in germany}.
\newblock \bibinfo{journal}{Energy} \bibinfo{volume}{158}, \bibinfo{pages}{1130--1149}.
\bibitem[{Yoon et~al.(2022)Yoon, Seo and Lee}]{Yoon2022}
\bibinfo{author}{Yoon, H.J.}, \bibinfo{author}{Seo, S.K.}, \bibinfo{author}{Lee, C.J.}, \bibinfo{year}{2022}.
\newblock \bibinfo{title}{Multi-period optimization of hydrogen supply chain utilizing natural gas pipelines and byproduct hydrogen}.
\newblock \bibinfo{journal}{Renewable and Sustainable Energy Reviews} \bibinfo{volume}{157}, \bibinfo{pages}{112083}.
\bibitem[{Yu et~al.(2024)Yu, Hao, Ali, Hua and Sun}]{Yu2024}
\bibinfo{author}{Yu, Q.}, \bibinfo{author}{Hao, Y.}, \bibinfo{author}{Ali, K.}, \bibinfo{author}{Hua, Q.}, \bibinfo{author}{Sun, L.}, \bibinfo{year}{2024}.
\newblock \bibinfo{title}{Techno-economic analysis of hydrogen pipeline network in china based on levelized cost of transportation}.
\newblock \bibinfo{journal}{Energy Conversion and Management} \bibinfo{volume}{301}, \bibinfo{pages}{118025}.
\bibitem[{Zhang et~al.(2024)Zhang, Jia, Bai, Wang, An, Zhao and Sun}]{Zhang2024}
\bibinfo{author}{Zhang, L.}, \bibinfo{author}{Jia, C.}, \bibinfo{author}{Bai, F.}, \bibinfo{author}{Wang, W.}, \bibinfo{author}{An, S.}, \bibinfo{author}{Zhao, K.}, \bibinfo{author}{Sun, H.}, \bibinfo{year}{2024}.
\newblock \bibinfo{title}{A comprehensive review of the promising clean energy carrier: Hydrogen production, transportation, storage, and utilization (hptsu) technologies}.
\newblock \bibinfo{journal}{Fuel} \bibinfo{volume}{355}, \bibinfo{pages}{129455}.
\bibitem[{Zhao(2022)}]{Zhao2022}
\bibinfo{author}{Zhao, G.}, \bibinfo{year}{2022}.
\newblock \bibinfo{title}{Path optimization of hydrogen supply chain based on the milp model}.
\newblock \bibinfo{journal}{Natural Gas Industry} \bibinfo{volume}{42}, \bibinfo{pages}{118--124}.

\end{thebibliography}

\appendix
\section*{Appendix A: Detailed Mathematical Formulations for Hydrogen Hub Integration}
\setcounter{equation}{0}
\renewcommand\theequation{A.\arabic{equation}}

\subsection*{1.\;Objective Functions}

\begin{equation}
  \min \mathrm{LCH}_2
  = \frac{\text{TC}}{\text{TV}}
  =\frac{\text{CC} + \text{MC} + \text{OC}}{\sum_{k,j,r,t} V_{kjrt}}
  \label{eqA2}
\end{equation}

\begin{equation}
  \text{CC}
  =  \mathbit{C}_{\mathbit{HC}}
  +  \mathbit{C}_{\mathbit{VC}}
\end{equation}

\begin{equation}
  \mathbit{C}_{\mathbit{HC}}
  = \sum_{t}\mathrm{PV}_{t}\!
    \left(
      \sum_{i,k} B\!H_{ikt}^{\mathrm{new}}\,C_{ikt}^{KH}\,L_{ik}
      +\sum_{k,j} B\!H_{kjt}^{\mathrm{new}}\,C_{kjt}^{KH}\,L_{kj}
    \right)
  \label{eqA5}
\end{equation}

\begin{equation}
  \mathbit{C}_{\mathbit{VC}}
  = \sum_{t}\mathrm{PV}_{t}
    \sum_{r} C_{rt}^{KV}\,N\!V_{rt}^{\mathrm{new}}
  \label{eqA6}
\end{equation}

\begin{equation}
  \text{MC}
  = \mathbit{C}_{\mathbit{FO}}
  + \mathbit{C}_{\mathbit{LO}}
  + \mathbit{C}_{\mathbit{HO}}
  \label{eqA7}
\end{equation}

\begin{equation}
  \mathbit{C}_{\mathbit{FO}}
  = \sum_{t}\mathrm{PV}_{t}
    \sum_{k,j}\sum_{r\neq1}
    P_{rt}^{\mathrm{fuel}}
    \frac{2\,L_{kj}\,V_{kjrt}}{F\!E_{r}\,V_{r}^{\mathrm{cap}}},
  \quad \forall r\neq1
  \label{eqA8}
\end{equation}

\begin{equation}
  \mathbit{C}_{\mathbit{LO}}
  = \sum_{t}\mathrm{PV}_{t}
    \sum_{k,j}\sum_{r\neq1}
    P_{rt}^{\mathrm{wage}}
    \frac{2\,L_{kj}/S\!P_{r}+L\!T_{r}}{V_{r}^{\mathrm{cap}}}
    V_{kjrt},
  \quad \forall r\neq1
  \label{eqA9}
\end{equation}

\begin{equation}
  \mathbit{C}_{\mathbit{HO}}
  = \sum_{t}\mathrm{PV}_{t}\!
    \left(
      \sum_{i,k} B\!H_{ikt}\,P_{t}^{H}\,L_{ik}
      +\sum_{k,j} B\!H_{kjt}\,P_{t}^{H}\,L_{kj}
    \right)
  \label{eqA10}
\end{equation}

\begin{equation}
  \text{OC}
  = \mathbit{C}_{\mathbit{HL}}
  + \mathbit{C}_{\mathbit{CL}}
  + \mathbit{C}_{\mathbit{SL}}
  \label{eqA11}
\end{equation}

\begin{equation}
  \mathbit{C}_{\mathbit{HL}}
  = \sum_{t}\mathrm{PV}_{t}\,\alpha
    \left(
      \sum_{i,k}
        w_{1}^{\mathrm{loss}}\,L_{ik}
        \frac{V_{ikrt}}{V_{1}^{\mathrm{cap}}}
      +\sum_{k,j,r}
        w_{r}^{\mathrm{loss}}\,L_{kj}
        \frac{V_{kjrt}}{V_{r}^{\mathrm{cap}}}
    \right)
  \label{eqA12}
\end{equation}

\begin{equation}
  \mathbit{C}_{\mathbit{CL}}
  = \sum_{t}\mathrm{PV}_{t}\,\delta
    \sum_{i,j}\sum_{r\neq1}
    e m_{r}\,L_{kj}\,
    \frac{N\!V_{rt}}{F\!E_{r}},
  \quad \forall r\neq1
  \label{eqA13}
\end{equation}

\begin{equation}
  \mathbit{C}_{\mathbit{SL}}
  = \sum_{t}\mathrm{PV}_{t}\,\lambda
    \sum_{j}\bigl(S_{jt}^{\mathrm{pos}}+S_{jt}^{\mathrm{neg}}\bigr)
  \label{eqA14}
\end{equation}

\subsection*{2.\;Constraints}

\subsubsection*{Mass Balance Constraints}

\noindent\textbf{Supply Capacity Limit:}\\
\begin{equation}
  \sum_{k} V_{ikt} \;\le\; Q_{it}^{S},
  \quad \forall i,t
  \label{eqA15}
\end{equation}

\noindent\textbf{Hub Balance Constraint:}\\
\begin{equation}
  \sum_{i} V_{ikt}
  \;=\;
  \sum_{j,r} V_{kjrt},
  \quad \forall k,t
  \label{eqA16}
\end{equation}

\noindent\textbf{Demand Satisfaction:}\\
\begin{equation}
  \sum_{k,r} V_{kjrt}
  - w_{r}^{\mathrm{loss}} L_{kj}
    \frac{V_{kjrt}}{V_{r}^{\mathrm{cap}}}
  = Q_{jt}^{D}+S_{jt},
  \quad \forall j,t
  \label{eqA17}
\end{equation}

\begin{equation}
  S_{jt}=S_{jt}^{\mathrm{pos}}-S_{jt}^{\mathrm{neg}},
  \quad \forall j,t
  \label{eqA18}
\end{equation}

\begin{equation}
  S_{jt}^{\mathrm{pos}}\ge 0,
  \quad \forall j,t
  \label{eqA19}
\end{equation}

\begin{equation}
  S_{jt}^{\mathrm{neg}}\ge 0,
  \quad \forall j,t
  \label{eqA20}
\end{equation}

\noindent\textbf{Cluster Assignment Constraint:}\\
\begin{equation}
  B\!H_{kjt}=0,\;\;
  V_{kjrt}=0,
  \quad \text{if } k\neq c_j,\;
  \forall k,j,r,t
  \label{eqA21}
\end{equation}

\subsubsection*{Pipeline Constraints}

\noindent\textbf{Pipeline Usage Restriction:}\\
\begin{equation}
  V_{ikt}\le M\,B\!H_{ikt},\quad
  V_{kjrt}\le M\,B\!H_{kjt},
  \quad r=1,\;\forall i,k,j,t
  \label{eqA22}
\end{equation}

\noindent\textbf{Pipeline Construction System Limit:}\\
\begin{equation}
  \sum_{i,k} B\!H_{ikt}^{\mathrm{new}}
  +\sum_{k,j} B\!H_{kjt}^{\mathrm{new}}
  \le \mathrm{NH}_{\max},
  \quad \forall t
  \label{eqA23}
\end{equation}

\noindent\textbf{ Pipeline Construction Dynamics:}\\
\noindent\textbf{1) Normal Case: $\mathrm{NH}_{\mathrm{gap}}>\mathbf{0}$}\\

\begin{equation}
  B\!H_{ik0}=0,\;\;
  B\!H_{kj0}=0,\quad \forall i,k,j
  \label{eqA24}
\end{equation}

\begin{equation}
  B\!H_{ik\tau}=B\!H_{ik,\tau-1},\;\;
  B\!H_{kj\tau}=B\!H_{kj,\tau-1},
  \quad \forall i,k,j,\;
  t\le \tau < t+\mathrm{NH}_{\mathrm{gap}}
  \label{eqA25}
\end{equation}

\begin{equation}
  B\!H_{ik,t+\mathrm{NH}_{\mathrm{gap}}}
   = B\!H_{ik,t+\mathrm{NH}_{\mathrm{gap}}-1}
     +B\!H_{ikt}^{\mathrm{new}},
  \quad
  B\!H_{kj,t+\mathrm{NH}_{\mathrm{gap}}}
   = B\!H_{kj,t+\mathrm{NH}_{\mathrm{gap}}-1}
     +B\!H_{kjt}^{\mathrm{new}},
  \quad \forall i,k,j,t
  \label{eqA26}
\end{equation}

\begin{equation}
  B\!H_{ikt}^{\mathrm{new}}=0,\;\;
  B\!H_{kjt}^{\mathrm{new}}=0,
  \quad
  \forall i,k,j,t
  \;\text{where } t+\mathrm{NH}_{\mathrm{gap}}>T
  \label{eqA27}
\end{equation}

\noindent\textbf{2) Special Case: $\mathrm{NH}_{\mathrm{gap}}=\mathbf{0}$}

\begin{equation}
  B\!H_{ikt}=
  \begin{cases}
    B\!H_{ik0}^{\mathrm{new}}, & t=0,\\[4pt]
    B\!H_{ik,t-1}+B\!H_{ikt}^{\mathrm{new}}, & t\ge 1,
  \end{cases}
  \quad
  B\!H_{kjt}=
  \begin{cases}
    B\!H_{kj0}^{\mathrm{new}}, & t=0,\\[4pt]
    B\!H_{kj,t-1}+B\!H_{kjt}^{\mathrm{new}}, & t\ge 1,
  \end{cases}
  \quad \forall i,k,j
  \label{eqA28}
\end{equation}

\noindent\textbf{Pipeline Lifespan Management:}\\
\begin{equation}
  B\!H_{ik,t+T_{r=1}^{\max}}
  \le 1-B\!H_{ikt}^{\mathrm{new}},\;\;
  B\!H_{kj,t+T_{r=1}^{\max}}
  \le 1-B\!H_{kjt}^{\mathrm{new}},
  \quad \forall i,k,j,t
  \;\text{where } t+T_{r=1}^{\max}\le T
  \label{eqA29}
\end{equation}

\noindent\textbf{Pipeline Flow Capacity:}\\
\begin{equation}
  V_{ikt}
  \le F_{\mathrm{base}}^{\max}\,
     \frac{L_{\mathrm{base}}^{H}}{L_{ik}},\;\;
  V_{kjrt}
  \le F_{\mathrm{base}}^{\max}\,
     \frac{L_{\mathrm{base}}^{H}}{L_{kj}},
  \quad r=1,\;\forall i,k,j,t
  \label{eqA30}
\end{equation}

\subsubsection*{Vehicle Constraints}

\noindent\textbf{Minimum Vehicle Requirement:}\\
\begin{equation}
  N\!V_{rt}\;\ge\;
  \sum_{k,j}
  \frac{V_{kjrt}}{A\!T_{r}\,V_{r}^{\mathrm{cap}}}
  \frac{2L_{kj}}{S\!P_{r}+L\!T_{r}}
  \cdot\frac{1}{365},
  \quad \forall r\neq1
  \label{eqA31}
\end{equation}

\noindent\textbf{Vehicle Purchase and Management:}\\

\begin{equation}
  N\!V_{r0}=N\!V_{r0}^{\mathrm{new}},
  \quad \forall r\neq1
  \label{eqA32}
\end{equation}

\begin{equation}
  N\!V_{rt}=N\!V_{rt}^{\mathrm{new}}
            +N\!V_{r,t-1}
            -N\!V_{rt}^{\mathrm{scrap}},
  \quad \forall t\ge1,\;r\neq1
  \label{eqA33}
\end{equation}

\begin{equation}
  N\!V_{rt}^{\mathrm{scrap}}=
  \begin{cases}
    N\!V_{r,t-T_{r}^{\max}}^{\mathrm{new}}, & t\ge T_{r}^{\max},\\[4pt]
    0, & t<T_{r}^{\max},
  \end{cases}
  \quad \forall r\neq1
  \label{eqA34}
\end{equation}

\subsubsection*{Environmental Constraint}

\begin{equation}
  \sum_{k,r}
  e m_{r}\,L_{kj}\,
  \frac{N\!V_{rt}}{F\!E_{r}}
  \;\le\; E\!M_{jt}^{\mathrm{ceil}},
  \quad \forall j,t
  \label{eqA35}
\end{equation}

\end{document}